\documentclass[12pt]{article}

\include{setup}

\begin{document}

\thispagestyle{empty}
\def\thefootnote{\fnsymbol{footnote}}
\setcounter{footnote}{1}
\null
\strut\hfill SFB/CPP-14-01\\
\strut\hfill TTP14-001\\
\strut\hfill LPN14-001\\
\strut\hfill KA-TP-01-14\\
\strut\hfill MCnet-14-01\\
\vskip 0cm
\vfill
\begin{center}
  {\Large \boldmath{\bf Vector-boson pair production 
 and electroweak  corrections  in \HERWIG~}
\par} \vskip 2.5em
{\large
{\sc Stefan Gieseke$^{1}$, Tobias Kasprzik$^{2}$, 
     Johann H.\ K\"uhn$^{2}$
}\\[2ex]
{\normalsize \it $^1$Karlsruhe Institute of Technology (KIT), Institut f\"ur Theoretische Physik (IThP)\\
D-76131 Karlsruhe, Germany}
\\[1ex]
{\normalsize \it $^2$ Karlsruhe Institute of Technology (KIT), Institut
  f\"ur Theoretische \mbox{Teilchenphysik (TTP)}\\
D-76131 Karlsruhe, Germany
}\\[2ex]
}
\par \vskip 1em
\end{center}\par
\vskip .0cm \vfill {\bf Abstract:} The detailed study of vector-boson pair
production processes at the LHC will lead to a better understanding
of electroweak physics. As pointed out before, a
consistent inclusion of higher-order electroweak effects in the analysis
of corresponding experimental data may be crucial to properly predict
the relevant phenomenological features of these important
reactions. Those contributions lead to dramatic distortions of
invariant-mass and angular distributions at high energies, but may also
significantly affect the cross section near threshold, as is the case
e.g.\ for Z-pairs. For this reason, we present an analysis of the
next-to-leading-order electroweak corrections to WW, WZ and ZZ
production at the LHC, taking into account mass effects as well as
leptonic decays. Hence, our predictions are valid in the whole kinematic
reach of the LHC and, moreover, respect the spin correlations of the
leptonic decay products at NLO accuracy. Starting from these fixed-order
results, a simple and straight-forward method is motivated to combine
the electroweak corrections with state-of-the-art Monte Carlo
predictions, focusing on a meaningful combination of higher-order
electroweak and QCD effects. To illustrate  our
approach, the electroweak corrections are implemented in the \HERWIG{}
generator, and their phenomenological effects within a QCD environment
are studied explicitly.
 \par

\par
\vskip 1cm
\noindent
January 2014    
\mla
\par
\null
\setcounter{page}{0}
\clearpage
\def\thefootnote{\arabic{footnote}}
\setcounter{footnote}{0}

\section{Introduction}\label{se:intro}

Vector-boson pair production processes play a central role in LHC
phenomenology. These processes are not only of great importance with
respect to background analyses in Standard-Model (SM) Higgs production,
they will also provide deeper insight into the structure of the
electroweak interaction at highest energies. This is particularly true
for the future high-luminosity run of the LHC at a center-of-mass (CM)
energy of 13 TeV, which will allow for an unprecedented accuracy in the
analysis of vector-boson interactions at the TeV scale.  Consequently,
theoretical prediction with high accuracy are needed for this important
 class of processes.

 Given the experimental accuracy already achieved by LHC experiments in
 the 7- and 8-TeV runs~\cite{Aad:2012awa, ATLAS:2012mec, Aad:2012twa,
   Chatrchyan:2013yaa, Chatrchyan:2013oev, Chatrchyan:2012sga}, at least
 the next-to-leading order (NLO) QCD corrections are mandatory for a
 robust prediction of $V$-boson pair production processes (for selected
 references see Refs.~\cite{Bierweiler:2012kw,Bierweiler:2013dja}). In
 addition, even first steps towards NNLO predictions of massive
 vector-boson pair production processes have been
 made~\cite{Gehrmann:2013cxs, Chachamis:2008yb}. (As far as photon-pair
 and Z$\gamma$ production at the LHC are concerned, the full strong
 two-loop corrections are known even fully
 differentially~\cite{Bern:2001df, Catani:2011qz, Grazzini:2013bna}.) In
 particular, approximate NNLO results for $\mathrm{W^+Z}$ and WW
 production have been provided for high-transverse-momentum
 observables~\cite{Campanario:2012fk,Campanario:2013wta}, as well as for
 WW production in the threshold limit~\cite{Dawson:2013lya}.

However, at energies of a few hundred GeV, also electroweak (EW) corrections
(and other related electroweak effects) are becoming more and more
important, and a lot of activity has taken place also in this field. In
particular, the interplay of EW corrections and anomalous couplings has
been investigated in Ref.~\cite{Accomando:2005xp}. The corresponding EW
corrections have been computed in Ref.~\cite{Accomando:2004de} in the
high-energy limit, including leptonic decays and off-shell
effects. Recently, also the full EW corrections to W-pair production,
also taking into account mass effects, have been evaluated for the
leptonic final state~\cite{Billoni:2013aba}. Leading two-loop effects at
high transverse momenta were evaluated in Ref.~\cite{Kuhn:2011mh} for
W-pairs. A detailed analysis of on-shell $V$-boson pair production
($V=\PW^\pm,\PZ\,$) and $\gamma\gamma$ production including EW
corrections has been provided in Refs.~\cite{Bierweiler:2012kw,
  Bierweiler:2013dja}, consistently including all mass
effects. Recently, a detailed review of NLO effects in pair production
of massive bosons has been provided, emphasizing the importance of
photon-induced contributions~\cite{Baglio:2013toa}.

Expecting first results for the full NNLO QCD corrections to W-pair
production in the near future~\cite{Chachamis:2013aya}, a natural next
step would be the combination of EW and QCD predictions at
$\mathcal{O}(\alpha_s\alpha)$ accuracy on a consistent theory basis, as
has been partially done for the Drell--Yan
process~\cite{Boughezal:2013cwa} already. However, these multiscale two-loop
calculations are beyond feasibility at present. Nevertheless, at least a
pragmatic prescription of combining QCD predictions with EW corrections
is desirable, aiming for a combination of EW precision with standard
Monte Carlo (MC) tools.

In this work we extend the above results in two ways. In
Sect.~\ref{se:EW} a first study of EW corrections to massive
vector-boson pair production processes is presented, including mass
effects as well as leptonic decays to allow for a realistic event
definition in the leptonic decay modes. We point out that our results
are not restricted to the high-energy regime but are also valid at
moderate energies of $\sim 200$ GeV, where already corrections of $\sim$
5--10\% may be observed. Applying these results, in a second step we
propose a straight-forward and simple strategy for the implementation of
EW corrections in any state-of-the-art MC generator
(Sect.~\ref{se:herwig}), relying on $K$-factors for unpolarized
two-by-two scattering and the assumption of factorization of EW and QCD
effects, as will be motivated in Sect.~\ref{se:qcdew}. Our method will
be tested against an alternative implementation method especially
tailored for the \HERWIG{}~\cite{Bahr:2008pv, Arnold:2012fq} MC
generator. In Sect.~\ref{se:herwigresults} selected numerical results
obtained within the \HERWIG{} setup will be discussed.

\section{Electroweak corrections}\label{se:EW}
As stated in the introduction, the complete evaluation of the combined
QCD and electroweak corrections to gauge-boson pair production of order
$\alpha\alpha_s$ is presently out of reach. As a first step we,
therefore, consider a factorized ansatz, where the kinematics of events
with additional QCD radiation is mapped to effective two-body
collisions, described by effective (squared) partonic energies and
energy transfers $\hat s$ and $\hat t$.  We assume that the bulk of the
QCD corrections arises from events with soft or collinear emission,
where the factorized ansatz is expected to work well. The EW corrections
are taken directly from the result of the one-loop calculation,
evaluated at the same kinematical point. This approach is expected to
fail for events with large transverse momenta of the gauge boson
pair recoiling against a jet with large transverse momentum. These
events, however, are of lesser relevance for the study of gauge-boson
dynamics and can be eliminated by suitable cuts, as discussed in
Sect.~\ref{se:herwig}.  The motivation of this approach will be
discussed in Sect.~\ref{se:qcdew}, and details of the implementation are
given in Sect.~\ref{se:herwig}.

In the present section we concentrate on the electroweak corrections and
motivate that indeed the bulk of the electroweak corrections can be
collected in a $K$-factor which is given as a function of $\hat s$ and
$\hat t$ only (Sect.~\ref{se:onshell}).  Photonic corrections (which
evidently lead to a more complicated kinematic situation) can be split
off such that the corresponding modifications of the electroweak
corrections are small. This aspect will be investigated in
Sect.~\ref{se:vvgeneral}.

A second simplification is introduced by applying a correction factor
which does not depend on the helicities of the gauge bosons. In
Sect.~\ref{se:4ldetails} we argue that this approximation still
preserves the proper angular distributions and correlations of the Z
and W decay products and investigate the phenomenological implications
of this approximation in detail.  Finally, the corrections as derived
for on-shell gauge boson production are applied for the cases where W
or Z are slightly off mass shell (In the case of Z bosons we also
include the amplitude where fermion pairs are produced through the
virtual photon). The quality of this approximation at lowest order will
also be discussed in Sect.~\ref{se:4ldetails}. First, however, let us
briefly recall the most important phenomenological features of various
EW effects in on-shell WW, WZ and ZZ production at the LHC.

\subsection{On-shell gauge-boson pair productions}
\label{se:onshell}
\begin{figure}
\begin{center}
\includegraphics[width = 0.9\textwidth]{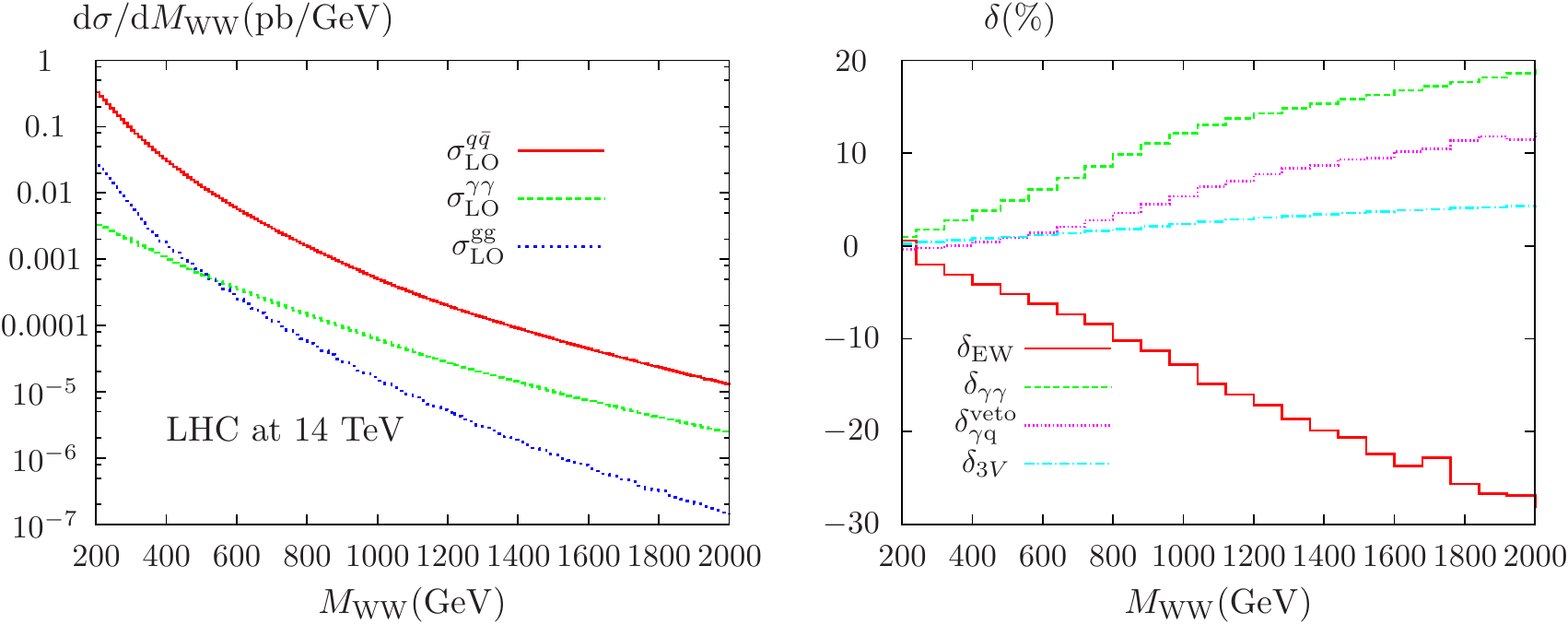}
\includegraphics[width = 0.9\textwidth]{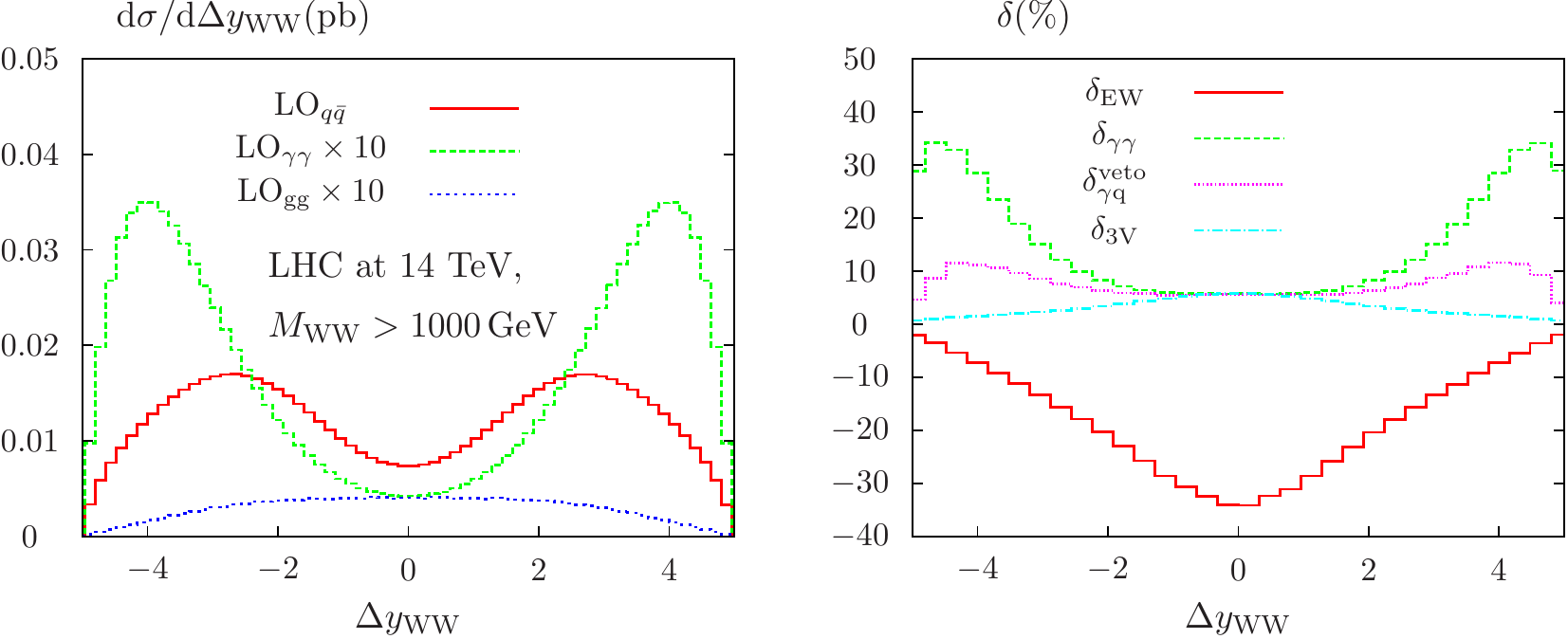}
\caption{Left: Differential LO cross sections for W-pair
  production at LHC14. Right: various EW corrections relative to the
  quark-induced LO process. Top: invariant-mass distribution; Bottom:
  WW rapidity-gap distribution for $ M_{\mathrm{WW}} > 1$ TeV. The results presented here
  are obtained in the default setup of
  Ref.~\cite{Bierweiler:2012kw}.}
\label{fi:WWonshell}
\end{center}
\end{figure}
\begin{figure}
\begin{center}
\includegraphics[width = 0.9\textwidth]{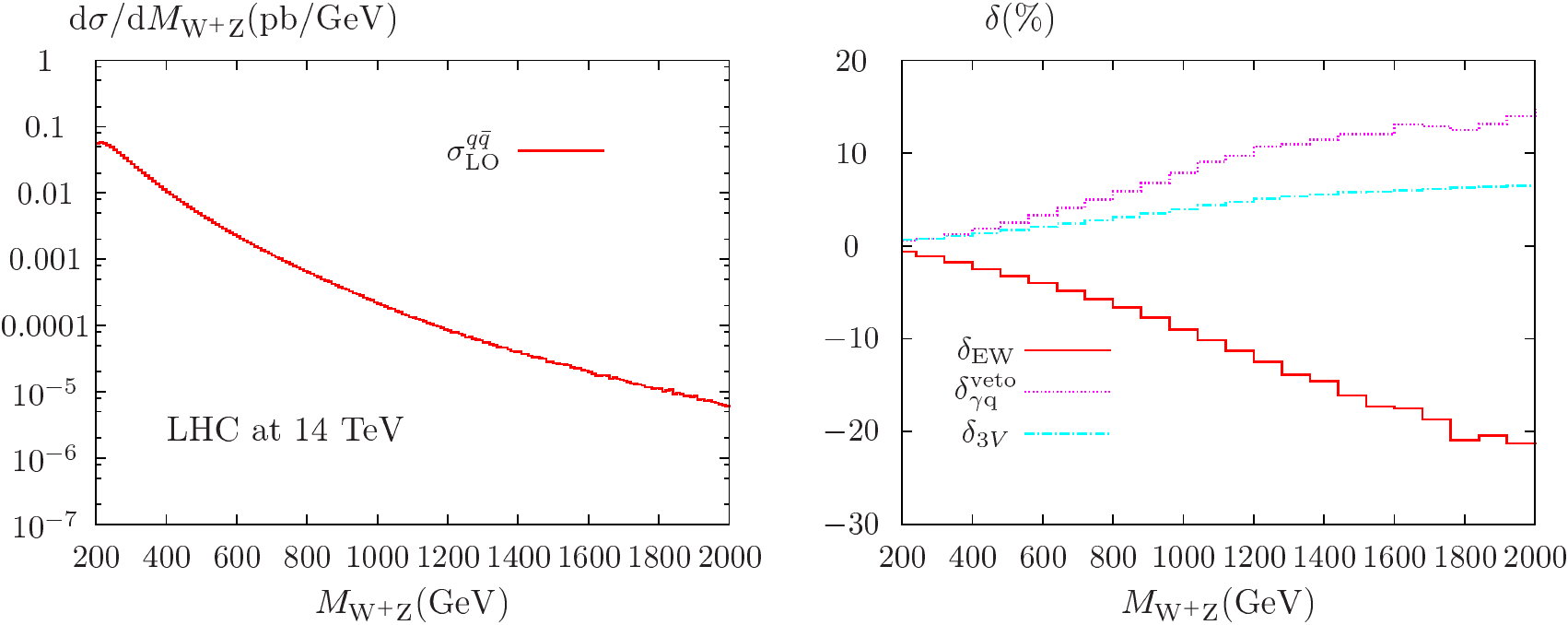}
\includegraphics[width = 0.9\textwidth]{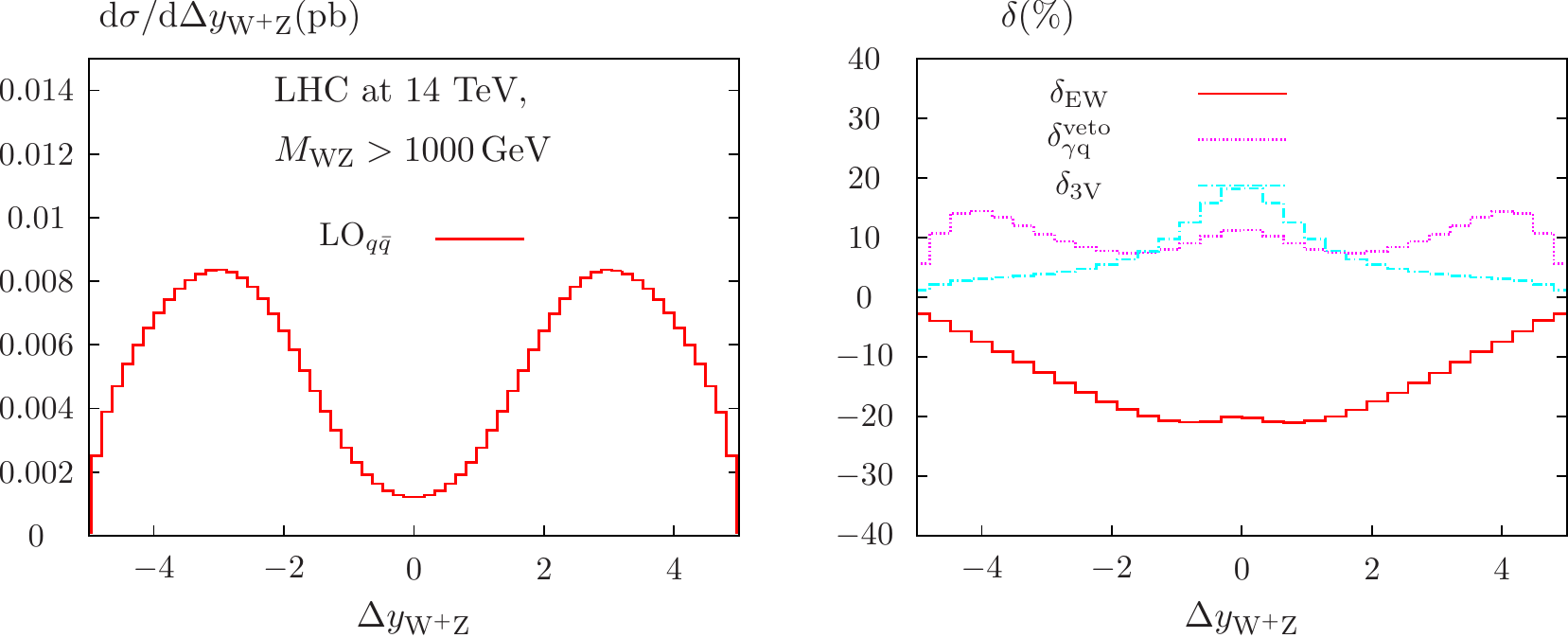}
\caption{Left: Differential LO cross sections for $\PW^+\PZ$
  production at LHC14. Right: various EW corrections relative to the
  quark-induced LO process. Top: invariant-mass distribution; Bottom:
  WZ rapidity-gap distribution for $ M_{\mathrm{WZ}} > 1$ TeV. The results presented here
  are obtained in the default setup of
  Ref.~\cite{Bierweiler:2013dja}.}
\label{fi:WZonshell}
\end{center}
\end{figure}
\begin{figure}
\begin{center}
\includegraphics[width = 0.9\textwidth]{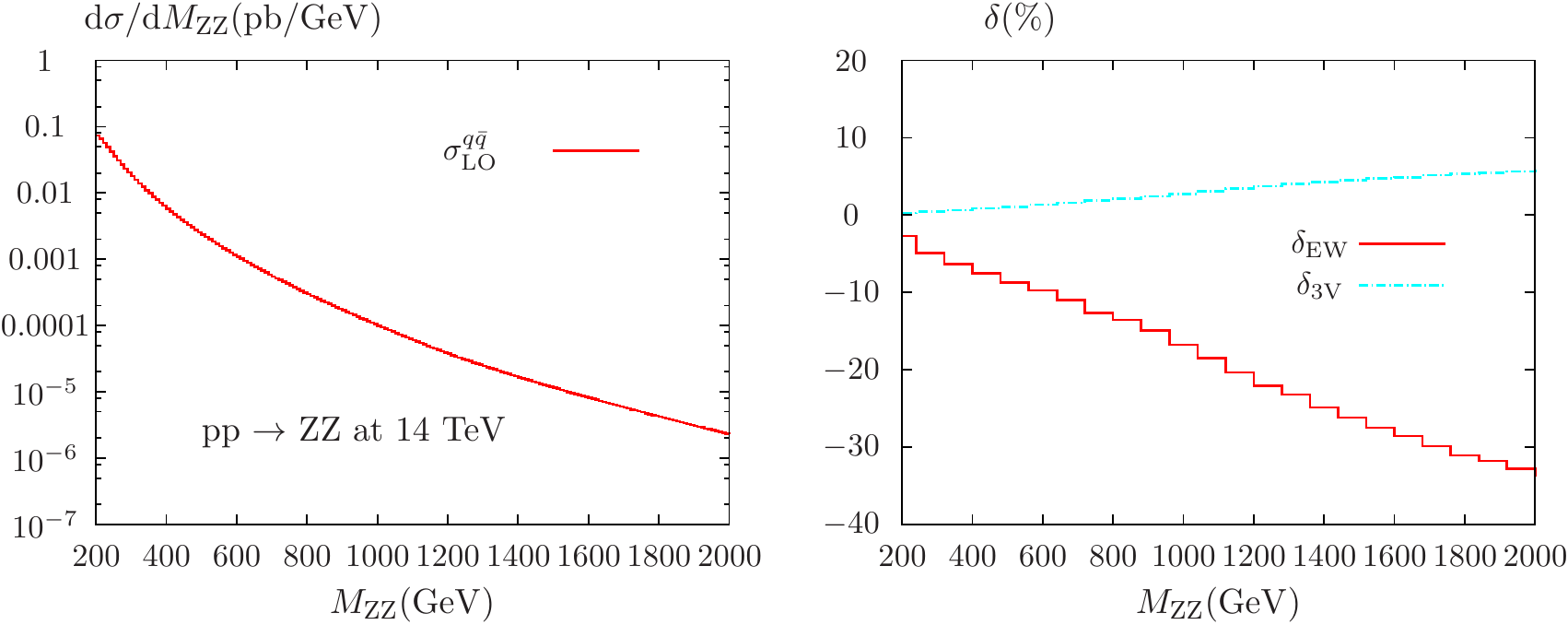}
\includegraphics[width = 0.9\textwidth]{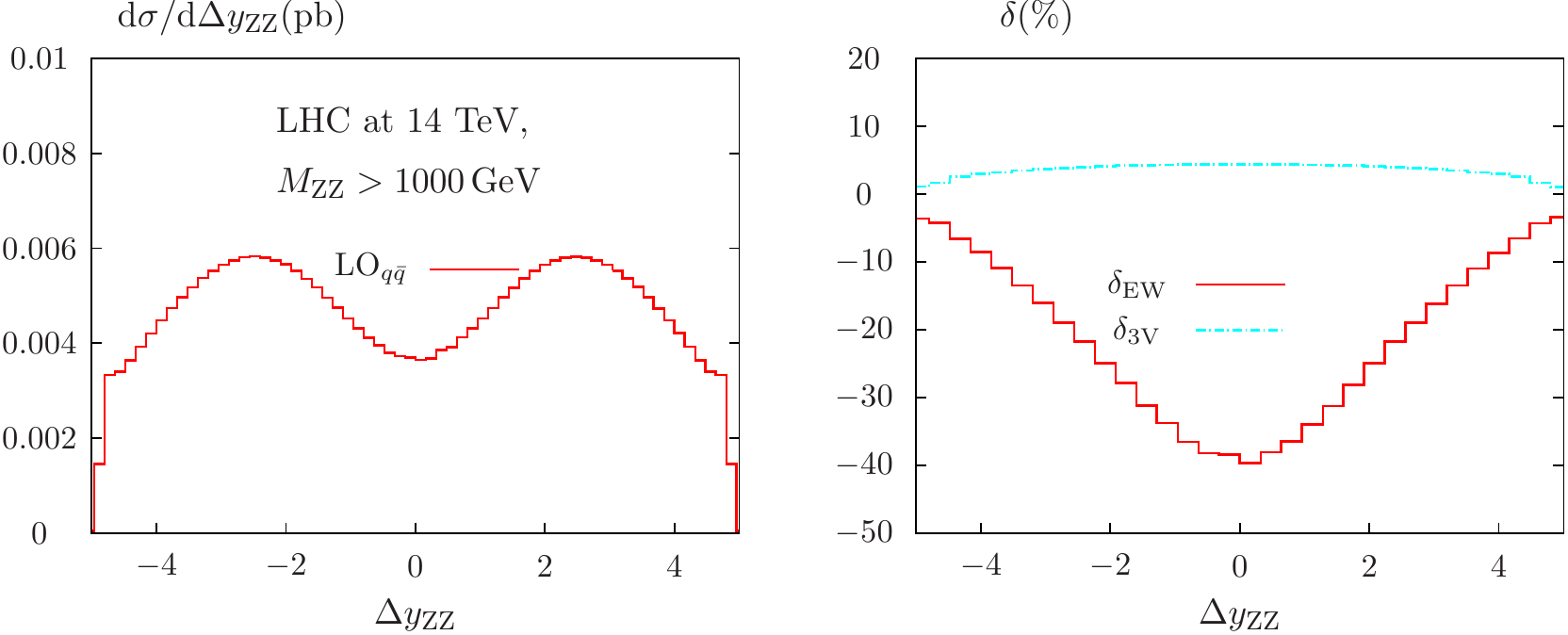}
\caption{Left: Differential LO cross sections for ZZ
  production at LHC14. Right: various EW corrections relative to the
  quark-induced LO process. Top: invariant-mass distribution; Bottom:
  ZZ rapidity-gap distribution for $ M_{\mathrm{ZZ}} > 1$ TeV. The results presented here
  are obtained in the default setup of
  Ref.~\cite{Bierweiler:2013dja}.}
\label{fi:ZZonshell}
\end{center}
\end{figure}
Here, we summarize the combination of different electroweak effects in
on-shell $V$-pair production which have been discussed in detail in
Refs.~\cite{Bierweiler:2012kw, Bierweiler:2013dja,Baglio:2013toa}. For
the numerical implementation, we use the default setup defined in
Refs.~\cite{Bierweiler:2012kw,Bierweiler:2013dja}.

In W-pair production, the invariant-mass distribution
(Fig.~\ref{fi:WWonshell} (top)) receives well-known logarithmically
enhanced negative EW corrections ($\delta_{\mathrm{EW}}$) growing with
energy. Positive contributions arise from the partonic subprocess
$\gamma\gamma \to \mathrm{WW}$ ($\delta_{\gamma\gamma}$) and the
photon-quark induced processes ($\delta_{\gamma q}^{\mathrm{veto}}$),
the latter evaluated applying the jet veto defined in
Ref.~\cite{Bierweiler:2012kw}. As already pointed out in
Ref.~\cite{Bierweiler:2013dja}, the effect of massive-boson radiation
($\delta_{3V}$) is moderate, however, strongly dependent on the event
selection.

The above picture significantly changes if angular distributions of the
W-pair are studied at high invariant masses. This can be seen in
Fig.~\ref{fi:WWonshell} (bottom) where distributions of the rapidity gap
of the two Ws are shown for $M_{\mathrm{WW}}>1000$ GeV. While the
genuine EW corrections drastically reduce the differential cross
section at high $p_{\rT,\mathrm{W}}$, corresponding to small rapidity
gap, the photon-induced contributions significantly increase the rate
at small scattering angles, corresponding to large rapidity gap. As a
result, a dramatic distortion of the angular distribution is visible which
might easily be misinterpreted as signal of anomalous couplings.

We point out that the photon-induced corrections presented above (which
are obtained using the MRST2004qed PDF set~\cite{Martin:2004dh} for the
photon density) suffer from a large systematic error stemming from our
ignorance of the photon content of the proton. This becomes obvious
looking at Fig.~25 of Ref.~\cite{Ball:2013hta}, where the
NNPDF2.3QED~\cite{Ball:2013hta} set has been used to estimate the error
on the $\gamma\gamma$-induced W-pair cross section. A relative error of
$\pm 50\%$ on the leading-order (LO) cross section at $M_{\mathrm{WW}} =
1000$ GeV can be deduced, solely induced by the photon PDF error. This
picture indicates that a significant improvement in the determination of
the photon PDFs is mandatory to reliably predict the W-pair production
cross section at high energies.

Turning to WZ production, the situation is qualitatively similar to WW
production, though here the $\gamma\gamma$ process is absent and the
genuine EW corrections are smaller (Fig.~\ref{fi:WZonshell}). In Z-pair
production, however, the $\gamma q$-induced contributions are
negligible~\cite{Baglio:2013toa}, and the real-radiation contributions
are always below 10\%. In total, particularly large negative
corrections, reaching $-40$\%, dramatically affect Z-pair production at
high invariant masses and transverse momenta (Fig.~\ref{fi:ZZonshell}).

\begin{figure}
\begin{center}
\includegraphics[width = 1.0\textwidth]{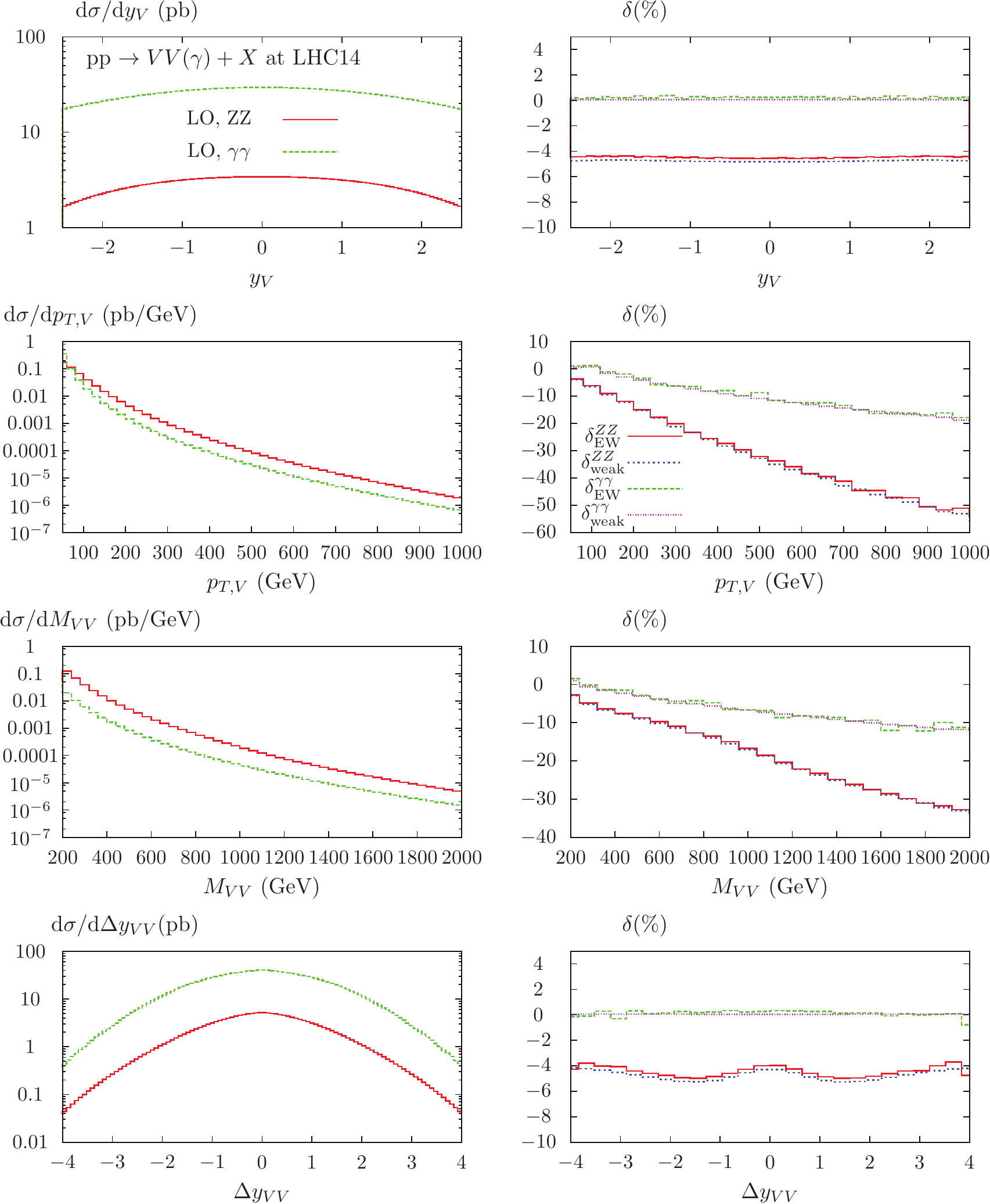}
\end{center}
\caption{\label{fi:zz_weak_os} Various differential distributions for
  on-shell Z-pair and $\gamma$-pair production at LHC14. Left: LO
  predictions; right: relative weak corrections
  ($\delta^{VV}_{\mathrm{weak}}$) and the full set of electroweak
  corrections ($\delta^{VV}_{\mathrm{EW}}$), including QED
  contributions. All results are obtained in the default setup defined
  in Ref.~\cite{Bierweiler:2013dja}.}
\end{figure}

\subsection{$K$-factors for the electroweak corrections}\label{se:vvgeneral}
Let us start with the simplest reaction, inclusive Z-boson pair
production.  The full set of electroweak corrections, including photon
radiation, Z-boson mass effects and virtual top quarks, has been
discussed in Ref.~\cite{Bierweiler:2013dja},  where the purely weak
  corrections were also evaluated for the 4-lepton final state in the
  pole approximation. The real and virtual QED corrections which can
be considered as gauge invariant subset were included in the on-shell
analysis.  However, their contribution is relatively small, in general
below 1\%. This is demonstrated in Fig.~\ref{fi:zz_weak_os} for four
characteristic distributions: the distributions in Z-boson rapidity
and transverse momentum, in the invariant mass of the Z pair and,
furthermore, in the difference between the rapidities of the two Z
bosons.  Note that in fact three correction functions are required, one
for the $\mathrm{u\bar u}$ induced reaction which we denote as
\begin{equation}
K^{\mathrm{ZZ}}_{\mathrm{u\bar u}}= 1+ \delta^{\mathrm{ZZ}}_{\mathrm{u\bar u}},
\end{equation}
the others, $K^{\mathrm{ZZ}}_{\mathrm{d\bar d}}$ and $K^{\mathrm{ZZ}}_{\mathrm{b\bar b}}$ for the $\mathrm{d\bar
d}$ and $\mathrm{b\bar b}$ induced reactions. All these factors depend on the
partonic $\hat s$ and $\hat t$ only. 

Similar considerations apply to di-photon production, and results
  analogous to ZZ production are also shown in
  Fig.~\ref{fi:zz_weak_os}. Again the relative contribution from purely
photonic corrections is small and the two-body approximation reproduces
the full answer to better than 1 percent.  The correction factors
$K^{\gamma\gamma}_{\mathrm{u\bar u}}$, $K^{\gamma\gamma}_{\mathrm{d\bar d}}$ and$
K^{\gamma\gamma}_{\mathrm{b\bar b}}$ are defined in an obvious way. The results presented in
Fig.~\ref{fi:zz_weak_os} were obtained using the default setup as defined in
Ref.~\cite{Bierweiler:2013dja}.

The situation is more involved in the case of W-pair and WZ
production where weak and electromagnetic corrections are intimately
intertwined. As is well known, virtual photonic corrections are
required to arrive at an ultraviolet finite answer.  The resulting
infrared divergencies are then canceled by real radiation, the
remaining collinear singularities are finally absorbed into a
redefinition of the PDFs.  Since one
is now dealing partially with a three body final state, the direct
use of a $K$ factor which depends on $\hat s$ and $\hat t$ only is
nontrivial.

In order to return to the kinematics of a two-to-two body reaction the
corrections from real radiation are now replaced by just subtracting
the endpoint singularities as defined originally in
Ref.~\cite{Dittmaier:1999mb} and described in the Appendix. As will be
shown explicitly in Sect.~\ref{se:4ldetails}, the physical
predictions remain practically unaffected by this simplification. (The
agreement at the level of around one percent is demonstrated
explicitly in Figs.~\ref{fi:wz_weak_4l} and~\ref{fi:ww_weak_4l}
through the comparison between $\delta_{\rm EW}^{\rm full}$ and
$\delta_{\rm EW}^{\rm V+E}$.) At the same time the kinematics of the resulting
final states is now fully described by the partonic $\hat s$ and $\hat
t$. In this way we obtain again three correction factors
$K^{\mathrm{WW}}_{\mathrm{u\bar u}}$, $K^{\mathrm{WW}}_{\mathrm{d\bar d}}$ and $K^{\mathrm{WW}}_{\mathrm{b\bar b}}$ which
can be employed in the framework of the Monte Carlo generator, just as
before. A similar approach is valid for W$^+$Z and W$^-$Z production,
which have, of course identical correction factors, denoted
$K^{\mathrm{WZ}}$. The corresponding endpoint contributions are also listed in
the Appendix.

\subsection{Gauge-boson polarization and four-lepton production}
\label{se:4ldetails}
Our corrections are presented for unpolarized gauge bosons, which in the
case of W and Z are observed through their decay products.  The
distributions of the decay products, however, are affected by the boson
polarization. Hence an additional complication could result from the
fact that the polarization pattern of the gauge bosons is modified by
the radiative corrections. In principle one would have to employ $K$
factors for the full set of helicity amplitudes. However, as
demonstrated in Ref.~\cite{Bierweiler:2013dja} for ZZ production, in
practice a fairly simple pattern emerges. Let us first consider the case
of Z pairs: For small transverse momenta the electroweak corrections are
small (about -4\%) and of similar magnitude for all four combinations of
transverse and longitudinal polarizations.  For large transverse momenta
one single configuration dominates completely and corrections for the
subdominant combinations are irrelevant. This feature has been
demonstrated in Table 7 of Ref.~\cite{Bierweiler:2013dja}, where the
cross sections and the corrections are displayed in the \mbox{low-,}
intermediate- and large-$p_\rT$ region, separated according to longitudinal
and transverse polarizations.

From these considerations it becomes clear that in case of Z-pair
production a single partonic $K$ factor is sufficient for large as well
as for small transverse momenta, and polarizations of the gauge bosons
and, correspondingly, the correlations between the decay products as
predicted in Born approximation are maintained even after inclusion of
the electroweak corrections.  The situation is slightly different for
W-pair (and also WZ) production. Here, roughly 10\% of the LO cross
section are given by longitudinally polarized Ws even at high transverse
momenta, and the corresponding relative EW corrections are substantially
different compared to the transversely-polarized case (see Fig.~7 of
Ref.~\cite{Kuhn:2011mh}). However, the numerical effects from the
longitudinal polarization on the radiative corrections are small, and it
is still sufficient -- as will be shown later -- to apply unpolarized
$K$-factors to reproduce the full corresponding EW corrections with
sufficient accuracy.

The corrections evaluated in Refs.~\cite{Bierweiler:2013dja,
  Bierweiler:2012kw} and encoded in our $K$ factors were obtained for
on-shell Z or W bosons. Any realistic simulation of four-fermion
production must, necessarily, include contributions from off-shell
configurations. In the case of Z also digrams with off-shell Z replaced
by virtual photons would be required for a description away from the Z
peak. However, for the experimental analysis of gauge boson production
the invariant mass of the decay products (lepton pairs or jets) must be
restricted to an interval around the nominal mass, say $|M_{l\bar
  l}-M_\PZ|<25\ {\rm GeV}$, to suppress the admixture of virtual photons
and enhance close-to-mass-shell gauge bosons. In this case the
neglection of virtual photons as implemented in \HERWIG{} can be
justified\footnote{The \HERWIG{} implementation of $V$-pair
  production~\cite{Hamilton:2010mb} relies on the double-pole
  approximation, where only doubly-resonant contributions are taken into
  account including off-shell effects.}.

In our approach off-shell effects and non-resonant contributions are
consistently accounted for in the LO predictions, while at NLO the NWA is
applied to compute the relative corrections. This approximation is well
motivated, since off-shell effects in the EW corrections only amount
to $\sim 0.5\%$, as demonstrated in Ref.~\cite{Billoni:2013aba}
for W-pair production.

From these considerations it can be expected that the dominant
weak corrections to four-lepton production at the LHC are well described
by process dependent $K$-factors which can be taken from the
unpolarized results for the corresponding $2 \to 2$ production
process. The validity of the approximations discussed above will now
be studied in detail.

\begin{figure}
\begin{center}
\includegraphics[width = 1.0\textwidth]{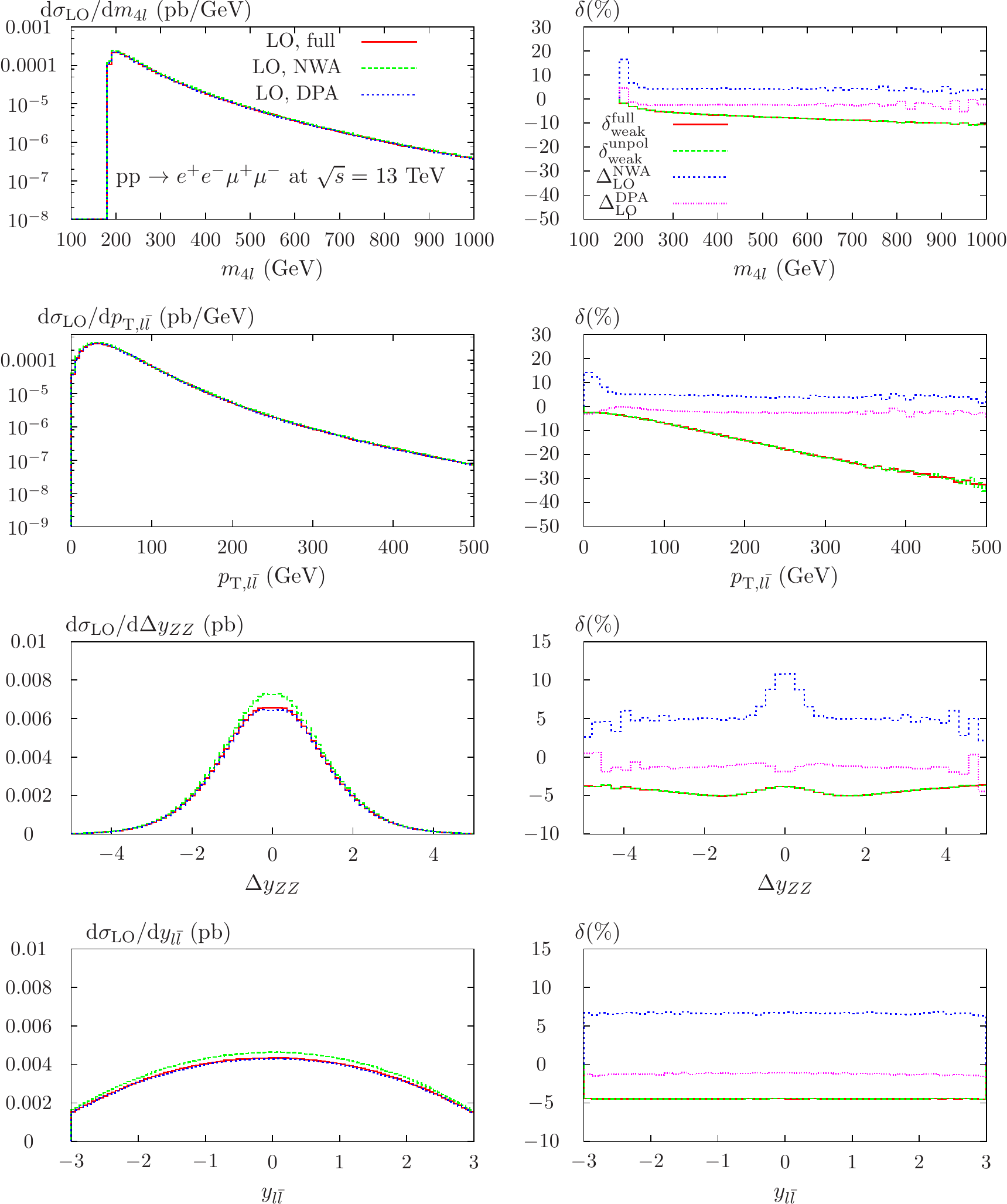}
\end{center}
\caption{\label{fi:zz_weak_4l} Various differential distributions for
  $e^+e^-\mu^+\mu^-$ production at LHC13. Left: The full LO prediction as
  well as NWA and DPA are shown; right: relative weak corrections
  $\delta_{\mathrm{weak}}^{\mathrm{full}}$ evaluated in the NWA, including spin
  correlations; weak corrections evaluated with unpolarized $2 \to 2 $
  $K$-factors ($\delta_{\mathrm{weak}}^{\mathrm{unpol}}$); relative deviations of
  NWA and DPA w.r.t.\ the full LO are also shown.}
\end{figure}
\begin{figure}
\begin{center}
\includegraphics[width = 1.0\textwidth]{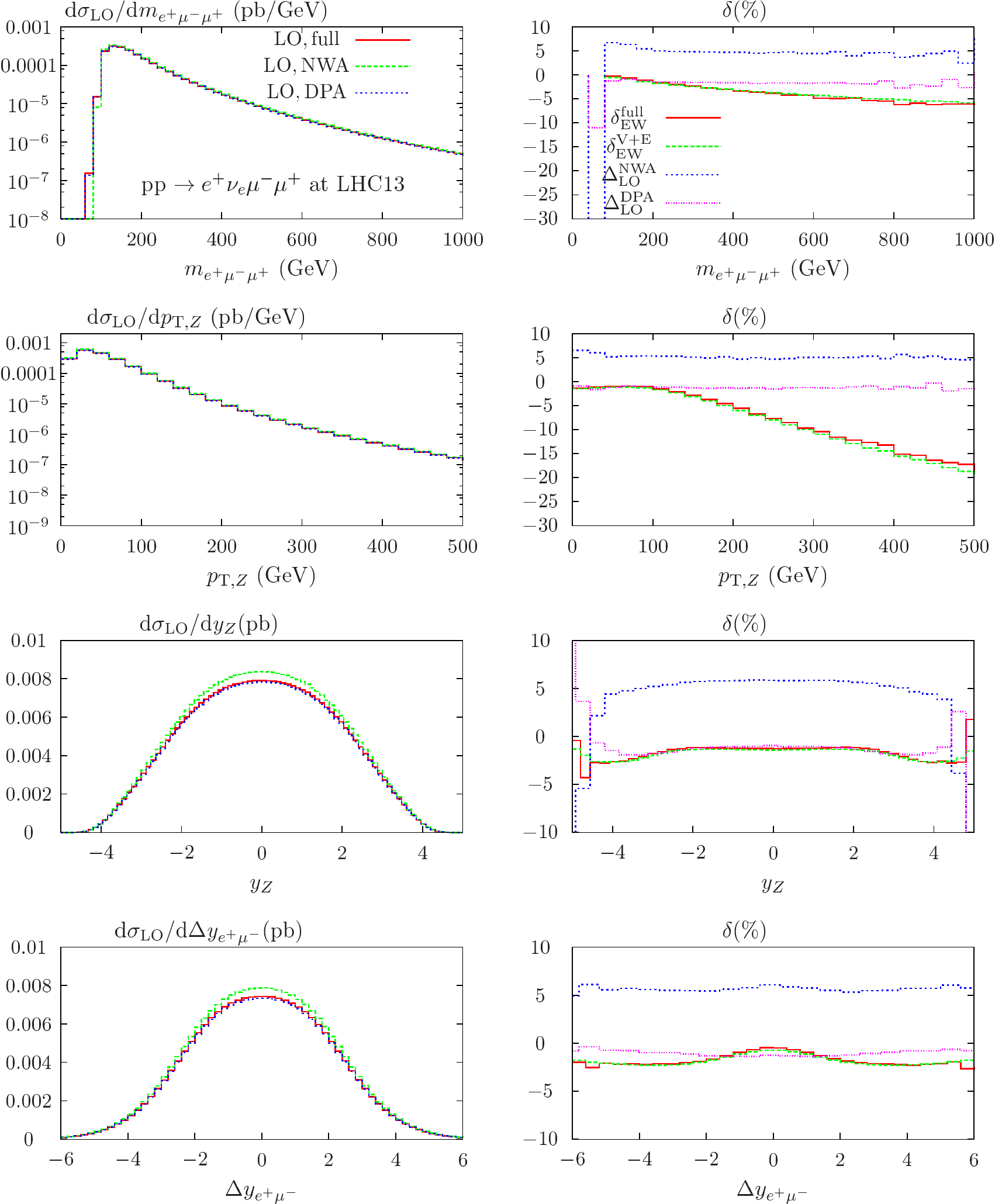}
\end{center}
\caption{\label{fi:wz_weak_4l} Various differential distributions for
  $e^+ \nu_e \mu^- \mu^+$ production at LHC13. Left: The full LO
  prediction as well as NWA and DPA are shown; right: full relative EW
  corrections $\delta_{\mathrm{EW}}^{\mathrm{full}}$ evaluated in the NWA including
  spin correlations; EW corrections evaluated with unpolarized $2 \to 2
  $ $K$-factors in the V+E approximation ($\delta_{\mathrm{EW}}^{\mathrm{V+E}}$);
  relative deviations of NWA and DPA w.r.t.\ the full LO are also shown.}
\end{figure}
\begin{figure}
\begin{center}
\includegraphics[width = 1.0\textwidth]{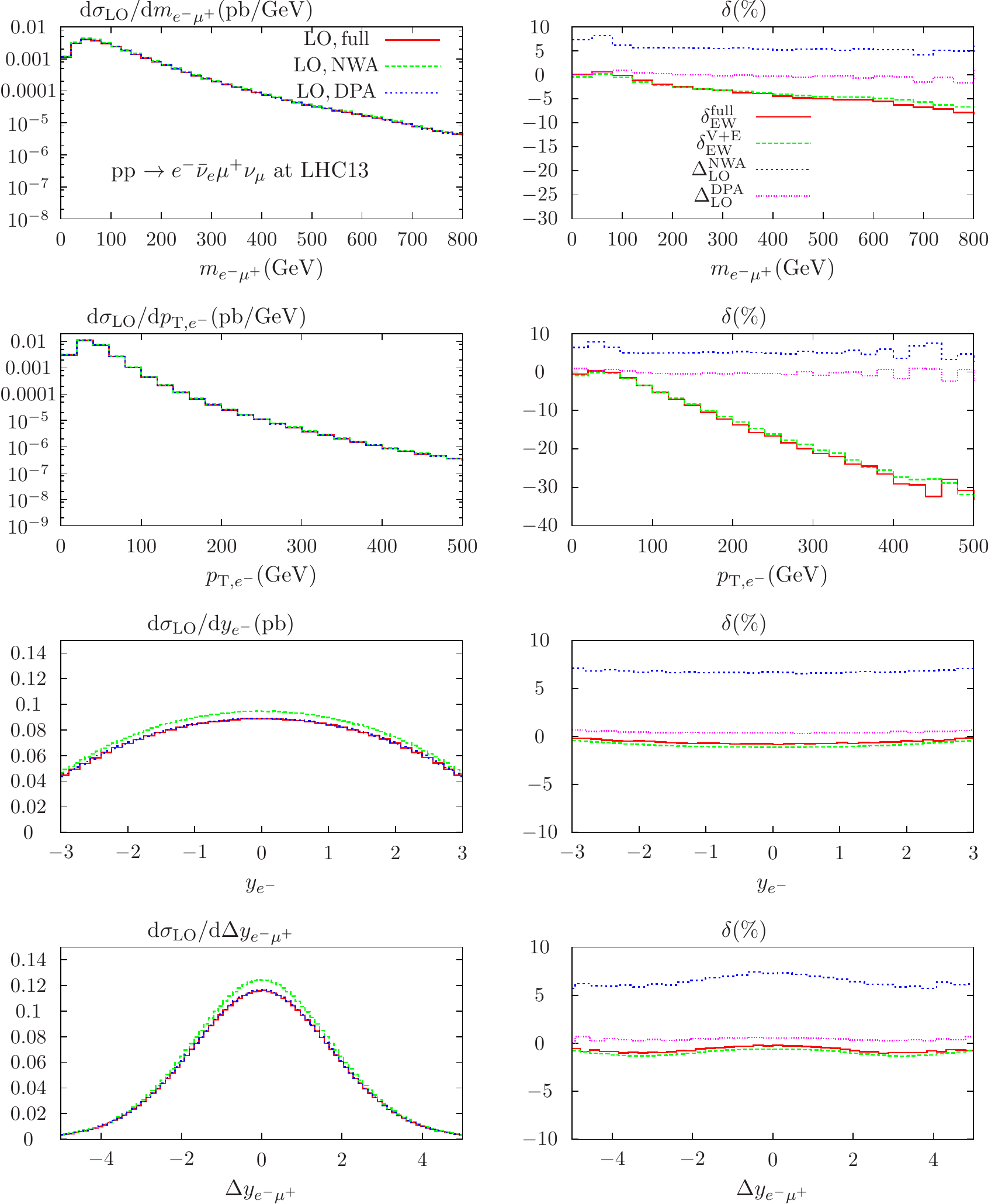}
\end{center}
\caption{\label{fi:ww_weak_4l} Various differential distributions for
  $e^- \bar{\nu}_e \mu^+ \nu_\mu$ production at LHC13. Left: The full LO
  prediction as well as NWA and DPA are shown; right: full relative EW
  corrections $\delta_{\mathrm{EW}}^{\mathrm{full}}$ evaluated in the NWA including
  spin correlations; EW corrections evaluated with unpolarized $2 \to 2
  $ $K$-factors in the V+E approximation ($\delta_{\mathrm{EW}}^{\mathrm{V+E}}$);
  relative deviations of NWA and DPA w.r.t.\ the full LO are also shown.}
\end{figure}
For this purpose we give a concise presentation of the computation of EW
corrections to massive gauge-boson pair production at the LHC,
consistently taking into account leptonic decays and, related to this,
spin correlations. In particular, we discuss the validity of the various
approximations discussed in the previous subsection.  To allow for a
sensible comparison with the \HERWIG~results presented in
Sect.~\ref{se:herwigresults}, we generally stick to the default
\HERWIG~setup for gauge-boson pair production as defined
in~Ref.\cite{Hamilton:2010mb}. Specifically, in the leptonic event
selection we apply the following basic cuts
\begin{equation}
p_{\rT,l} > 10 \GeV\,,\qquad |y_{l}| < 5
\end{equation}
for the charged-lepton transverse momenta and rapidities. If neutrinos
are present in the final state, a minimal missing transverse momentum of
\begin{equation}
p_{\mathrm{T,miss}}> 25\GeV
\end{equation}
is also required.  The invariant mass of the lepton pair
is restricted to
\begin{equation}
|M_{l\bar{l}}-M_V| < 25 \GeV
\end{equation} 
to suppress non-resonant backgrounds. For the gauge boson widths we use
the values
\begin{equation}
\Gamma_\PW =2.141\GeV \,,\qquad \Gamma_\PZ =2.4952\GeV \,,
\end{equation}
and the weak coupling constant is defined in the $G_\mu$ scheme to
systematically absorb universal corrections related to the running of
$\alpha$ to the weak scale in the LO predictions. All remaining SM input
parameters are directly carried over from
Ref.~\cite{Bierweiler:2013dja}.

For the evaluation of hadronic cross sections we use the CT10NLO PDF
set~\cite{Lai:2010vv} in the LHAPDF framework~\cite{Whalley:2005nh}, and
the CKM dependence in the WZ production channels is included at leading
order, while in the computation of EW corrections the CKM matrix is set
to unity.

At leading order we present full results, including non-resonant and
off-shell effects, as well as results in two different
approximations. As far as the full LO cross sections are concerned, we
have checked that the difference between a naive fixed width
implementation and results obtained in the Complex-Mass Scheme
(CMS)~\cite{Denner:2005fg} is at the per-mill level and hardly
visible. All results presented here for the full LO cross sections
therefore correspond to the naive fixed-width implementation. 

In addition to the full results we provide the results for V-boson pair
production in the double-pole approximation (DPA) originally
discussed in Ref.~\cite{Accomando:2004de}. Here, the amplitudes for V-pair
production and decays are evaluated on-shell, but the Breit--Wigner
shape of the resonance is included in the evaluation of the squared
matrix elements to account for the dominant off-shell effects. We apply
the on-shell projection procedure proposed in Ref.~\cite{Denner:2000bj}
to construct proper on-shell momenta of the intermediate bosons from the
four-particle phase-space. Note that in addition to the physical cuts
displayed above we impose a technical cut, $m_{4l} > M_{V_1}+M_{V_2}$,
on the 4-lepton invariant mass since the on-shell projection suggested
in Ref.~\cite{Denner:2000bj} only gives sensible results above
threshold. 

As a third alternative, we work in the narrow-width approximation
(NWA), where the gauge bosons are strictly forced on-shell from the
beginning via the replacement
\begin{equation}
\frac{1}{(p^2-M_V^2)^2 + M_V^2 \Gamma_V^2} \to \frac{\pi}{M_V\Gamma_V}\,\delta(p^2-M_V^2)
\end{equation}
for the resonant squared propagators. 

In Figs.~\ref{fi:zz_weak_4l}, \ref{fi:wz_weak_4l} and
\ref{fi:ww_weak_4l} we present results for WW, W$^+$Z and ZZ production
at the LHC. Since in our approach the EW corrections to the partonic
subprocesses are insensitive to the leptonic decay mode, in each case we
concentrate on one specific decay channel, namely
\begin{subequations}\label{eq:procLOpp4l}
\begin{eqnarray}
q\bar{q} &\to& (\PZ/\gamma^*)\,(\PZ/\gamma^*) \to \Pep \Pem \,\mu^+\mu^-\,,\\
u_i\bar{d}_j &\to& \PW^+\,(Z/\gamma^*) \to \Pep \nu_{\Pe} \,\mu^-\mu^+\,,\\
q\bar{q} &\to& \PW^-\,\PW^+ \to \Pem \bar{\nu}_{\Pe} \,\mu^+\nu_{\mu},.
\end{eqnarray}
\end{subequations}
Note that if intermediate Z bosons are present in the process, the
$\gamma^*$ contributions and all related interference contributions are
taken into account in the full LO results, while those contributions
are absent in DPA and NWA, respectively.  
  
In the left panels of Figs.~\ref{fi:zz_weak_4l}, \ref{fi:wz_weak_4l} and
\ref{fi:ww_weak_4l} we present LO results for various typical
differential distributions for processes \refeq{eq:procLOpp4l} at LHC13,
resepctively.  Besides the full results, the respective approximate
results in NWA and DPA are also shown, always taken in leading order. 

The right-hand-side panels of the respective plots show the relative
deviations of the NWA ($\Delta_{\mathrm{LO}}^{\mathrm{NWA}}$) and DPA
($\Delta_{\mathrm{LO}}^{\mathrm{DPA}}$) from the full results. For all
pair-production channels one observes that the NWA overshoots the full
LO predictions at the level of 5\%, while the DPA results give a good
approximation valid at the 2--3\% level. From the upper right plot of
Fig.~\ref{fi:zz_weak_4l} it becomes obvious that both approximate
predictions become crude near the Z-pair production threshold where
off-shell effects apparently become more important.

Let us now turn to the EW corrections. As stated before, we compute
the full EW corrections to the respective polarized vector-boson pair
production processes, and include the spin correlations in the decays
to leptons, which are treated at leading order. The
actual computation is carried out in the well-established {\tt
  Feyn\-Arts/Form\-Calc/Loop\-Tools}~\cite{Kublbeck:1990xc,
  Hahn:1998yk, Hahn:2000kx, Hahn:2001rv, vanOldenborgh:1989wn} setup
already used in Refs.\cite{Bierweiler:2013dja,Bierweiler:2012kw}, and
Madgraph~\cite{Alwall:2007st} was useful for internal checks. We do
not take into account the EW corrections to the leptonic decay
processes for various reasons. On the one hand, the bulk of the EW
corrections to Z and W decay are given by final-state photon radiation
(FSR), which leads to significant
distortions of the phase-space distributions of the leptonic decay
products. These FSR corrections, however, are
included in \HERWIG{}~\cite{Hamilton:2006xz} in the YFS
framework~\cite{Yennie:1961ad} and will therefore not be considered
here. On the other hand, electroweak corrections to the inclusive boson
decay widths are implicitly included in the experimentally determined
values for the branching ratios used in the \HERWIG{} framework. 
Additionally, we strictly
stick to the NWA for the computation of the EW corrections. In this
simplified approach, the corrections completely factorize into
corrections either to the production or the decay process, and no
non-factorizable corrections, which connect production and decay, have
to be considered. Those contributions have to be taken into account using
the DPA as demonstrated in Ref.~\cite{Accomando:2004de}.  However, it is
known that the non-factorizable corrections largely cancel in
sufficiently inclusive observables~\cite{Falgari:2013gwa}.

In addition to the full EW corrections
$\delta_{\mathrm{EW}}^\mathrm{full}$ (which contain proper spin
correlations and, in case of WW and WZ production, also the full set of
QED corrections to the respective production process) approximate
results $\delta_{\mathrm{EW}}^{\mathrm{V+E}}$ in the virtual+endpoint
(V+E) approximation are presented in the same plots, where unpolarized
on-shell $K$-factors have been used to obtain the relative corrections,
as detailed in Sect.~\ref{se:vvgeneral}. One observes that the
approximate ansatz gives an almost perfect approximation for the full
result, in general better than 1\%. The best agreement is observed for
ZZ production, while for WW production a slight discrepancy is
visible. This can be understood recalling that photon radiation and
related QED corrections are largest for WW production, as has been
demonstrated in Ref.~\cite{Bierweiler:2013dja}, while they remain small
in case of Z pairs. In general, the agreement between the full result
and the V+E approximation is even better than expected. As a conclusion
one finds that it is justified to use unpolarized $K$-factors in the V+E
approximation to describe vector-boson pair production at the LHC at a
sufficient accuracy.

\section{Electroweak and QCD corrections -- combined}
\label{se:qcdew}
As stated in the introduction, the complete treatment of combined QCD
and electroweak corrections, involving e.g.\ two-loop terms of order
$\alpha_s\alpha_{\rm weak}$ is presently out of reach. Consequently
neither an additive treatment of the corrections, 
$(1+\delta_{\rm QCD}+\delta_{\rm weak})$, nor a multiplicative one,
$(1+\delta_{\rm QCD})(1+\delta_{\rm weak})$, will lead to the correct
result. However, the bulk of the QCD corrections for the processes under
consideration is related to soft or collinear gluon radiation 
from the incoming quark-antiquark system. Soft radiation leaves the 
initial state practically unchanged, the quarks remain close to their
mass shell and the weak corrections can
be taken from the $\hat s$ and $\hat t$  dependent $K$ factor. 

A similar line of reasoning is applicable to hard collinear radiation
which leads to final states with vanishing or small transverse momentum
of the diboson system.
Let us assume that a collinear gluon is radiated from the incoming
(anti-)quark, which subsequently initiates the boson pair production. Also in
this case the (anti-)quark stays close to its mass shell. The scattering angle
of the $q_1\bar q_2\to V_1V_2$ reaction in the $V_1V_2$ rest frame can
again be directly identified with the scattering angle of the partonic
reaction. A similar line of reasoning applies to reactions with quarks
or antiquarks originating from gluon splitting.

The situation becomes more involved in the case of hard non-collinear
radiation, which leads to a diboson system of large transverse
momentum. In this case only an approximate prescription can be
formulated, since the original $2\to2$ kinematics is distorted. However,
this approximate prescription must coincide with the previous one in the
limit of vanishing transverse momentum.  To be specific we advocate the
following strategy to compute the effective partonic Mandelstam
variables $\hat{s}'$ and ${\hat t}'$ from the distorted kinematics for
the evaluation of $K({\hat s}', {\hat t}')$: The squared CM energy is
calculated from the four-lepton final state via
\begin{equation}\label{eq:m4l}
{\hat s}' = m_{4l}^2\,.
\end{equation}
The momenta are boosted into the four-lepton CM frame (denoted by
$\Sigma^*$). In this frame the unit directions of initial-state hadrons
shall be denoted by
\begin{equation}\label{eq:axis}
\vec{e}^{\,*}_i = \frac{\vec{p}^{\,*}_i}{|\vec{p}^{\,*}_i|}\,,\quad i=1,2\,.
\end{equation} 
The direction of the effective scattering axis in $\Sigma^*$ is now
defined by
\begin{equation}
\hat{\vec{e}}^{\,*} = \frac{\vec{e}^{\,*}_1 - \vec{e}^{\,*}_2}{|\vec{e}^{\,*}_1-\vec{e}^{\,*}_2|}\,,
\end{equation}
and the effective scattering angle is, correspondingly, given by
\begin{equation}\label{eq:costheta}
\cos {\theta}^* = \vec{v}^{\,*}_1\cdot \hat{\vec{e}}^{\,*}\,,
\end{equation}
where the $\vec{v}^{\,*}_i$ denotes the momentum direction of vector
boson $V_i$. The Mandelstam variable $\hat{t}'$ is then computed
  from $\theta^*$ assuming on-shell kinematics.   

Diboson events with large transverse momenta necessarily require the
presence of at least one hard quark or gluon jet, and electroweak
corrections would have to be evaluated separately for this class of
processes. As long, as they can be treated as a small admixture to the
diboson sample, suppressed by an additional factor $\alpha_s$, the
distortion of the weak corrections should not lead to a significant
error for the inclusive sample. As long as one is interested
specifically in the analysis of the diboson process, a cut on the
transverse momentum of the dibosons system will eliminate the
pollution with events of a very different nature. Let us discuss this
important issue in some more detail. As pointed out by several
groups~\cite{Bierweiler:2012kw, Campanario:2012fk}, at large
transverse momenta $V$-pair production is dominated by new topologies
which are absent at lowest order in QCD. These topologies
correspond to $V + \mathrm{jet}$ production with the radiation of an
additional $V$ from the quark jet rather than being a
correction to $V$-pair production, and  spoil the perturbative series for
the prediction of leptonic observables at high transverse momenta. They
lead to huge QCD $K$-factors together with large residual scale
uncertainties. To improve the corresponding theory predictions the
authors of Ref.~\cite{Campanario:2012fk} have provided approximate NNLO QCD
predictions for these particular observables in WZ production,
applying the {\tt LoopSim} method~\cite{Rubin:2010xp}. They observe
pronounced shifts of the predictions going from NLO to NNLO and at the
same time a significant reduction of the theory uncertainties.  In
this work, however, we follow another strategy to stabilize the
problematic high-$p_\rT$ observables, employing suitable restrictions
on the transverse momentum of the 4-lepton system, as will be
detailed in Sect.~\ref{se:herwigresults}.

In the present section a fairly general prescription has been
suggested how to implement weak corrections into a generic Monte Carlo
program which includes QCD corrections and hadronisation
already. Indeed this prescription allows for an {\it a posteriori}
application of EW corrections to any QCD Monte Carlo, at least as long
as the partonic origin ($\mathrm{u\bar{u}}$ vs $\mathrm{d\bar{d}}$) of the final state
remains under control.  In the next section a slightly different
approach will be described which is specifically tailored to the
program \HERWIG. In this case a partonic $\hat s$ and $\hat t$ is
introduced in the program from the very beginning and, in the limit of
diboson events with small transverse momenta, coincides with the
prescription described above. In the following section this
correspondence will be investigated in more detail.

\section{Implementation into \HERWIG~}\label{se:herwig}

Our starting point of the implementation of the EW corrections in
\HERWIG{} is the electroweak
$K$-factor.  Following the strategy of multiplicative QCD and EW
corrections we reweight the events that have been generated in \HERWIG{}
as the hard processes.  For vector-boson pair production \HERWIG{}
delivers unweighted events, hence we can compute $K(\hat s, \hat t)$ and use
this directly as a reweighting factor for each event such that $K(\hat
s, \hat t) < 1$ leads to a suppression of events with  given $(\hat s,
\hat t)$ while $K(\hat s, \hat t) > 1$ leads to an enhancement.  The
actual implementation of a reweighting factor for a given hard event is
straightforward in \HERWIG{} and \texttt{ThePEG} once the variables
$(\hat s, \hat t)$ are known.

Let us discuss the kinematics of our events in more detail.  For EW
corrections to the Born process the calculation of a $K$ factor is
straightforward as we can directly access the kinematic setup of the
hard process once this is generated.  In this case $(\hat s, \hat t)$
can be computed uniquely.  As soon as we want to apply the EW
corrections to an event that is generated from an NLO QCD matrix element
this is no longer the case.  In this case we face the complication that an event
with real radiation is not described by $2\to 2$ kinematics anymore.

Fortunately, the situation in the \HERWIG{} case is largely simplified
by the fact that the NLO QCD corrections are applied in the POWHEG
scheme \cite{Frixione:2007vw,Hamilton:2010mb}. (See
\cite{Buckley:2011ms,Gieseke:2013eva} for more details.)  The POWHEG
scheme provides a consistent matching of NLO corrections and parton shower
emissions without double counting.  In this scheme, the differential
cross section for the hard process can be written as
\begin{equation}
  \label{eq:dsigpowheg}
  \mathrm{d}\sigma = \bar B(\Phi_B)\,\mathrm{d}\Phi_B
  \left[\Delta_R(k_{\perp, {\rm min}}) 
    + \frac{R(\Phi_B, \Phi_R)}{B(\Phi_B)}
    \Delta_R(k_{\perp}(\Phi_B, \Phi_R))\,\mathrm{d}\Phi_R 
  \right]\ .
\end{equation}
Here, $\Phi_B$ and $\Phi_R$ denote the Born and radiative phase space
variables, respectively, and $k_{\perp, {\rm min}}$ is the minimum
transverse momentum that is generated by the parton shower.  $\Phi_R$
only parametrizes the additional variables to specify a hard emission
relative to the Born configuration.  $B(\Phi_B)$ and $R(\Phi_R)$ are the
Born and real-emission matrix elements squared for the hard process
under consideration.  $B(\Phi_B)$ is the Born differential cross section
while
\begin{equation}
  \label{eq:bbar}
  \bar B(\Phi_B) = B(\Phi_B) + V(\Phi_B) + \int R_S(\Phi_B, \Phi_R)\,
  \mathrm{d}\Phi_R\ .
\end{equation}
Here, $V(\Phi_B)$ is the (infrared finite) sum of virtual corrections
and the divergent part of the real corrections, while $R_S(\Phi_B,
\Phi_R)$ is the (also finite) real correction matrix element squared
with the divergent terms subtracted.  Finally, the POWHEG Sudakov form
factor is given by 
\begin{equation}
  \label{eq:powhegsud}
  \Delta_R(p_\perp) = \exp\left[
  -\int \mathrm{d}\Phi_R \,\frac{R(\Phi_B,
    \Phi_R)}{B(\Phi_B)}\Theta(k_\perp(\Phi_B, \Phi_R) - p_\perp)\right]\
 , 
\end{equation}
which, opposed to the Sudakov form factor in a common parton shower,
contains the full real emission matrix element squared.  The two terms
in Eq.~(\ref{eq:dsigpowheg}) are constructed to resemble the result of a
single parton shower emission applied to a configuration $\Phi_B$ with
weight $\bar B(\Phi_B)$.  The first term gives the no-emission
probability, the second the contribution from a single emission.
At the same time, upon expansion in $\alpha_s$,
Eq.~(\ref{eq:dsigpowheg}) reproduces the differential cross section at
NLO QCD.  

In the \HERWIG{} implementation we find exactly this prescription.  In a
first step we generate an event with kinematical configuration $\Phi_B$.
Technically, this is already the hard process.  Only in a second step,
already as part of the parton shower algorithm, the potential hard
emission with relative kinematics $\Phi_R$ is generated according to the
Sudakov form factor Eq.~(\ref{eq:powhegsud}).  Once this is done, the
default parton shower is used as at this point, all terms will become
formally higher than next-to-leading order and the default parton shower
is computationally much simpler than the Sudakov form factor
(\ref{eq:powhegsud}).

As the hard emission from POWHEG is technically already part of the
parton shower, we have direct access to the Born type variables $\Phi_B$
in the hard subprocess and hence we may easily compute the kinematic
variables $(\hat s, \hat t)$ for every event.  In effect, this assumes
that the EW corrections are applied only on the level of Born-type
kinematics and are not strongly influenced by the hard emission.  In
fact, we later also focus our analysis to regions where the transverse
momenta generated by hard gluon emissions are not too large by applying
a suitable veto.  This veto will suppress events where gauge-boson pairs
are accompanied by additional hard quark or gluon jets, leaving the
$q\bar{q}$ events largely unaffected. This also enforces the 
kinematics to be reasonably close to a Born configuration in order to
justify our approach.

Events with strong QCD activity, e.g.\ jets with large transverse
momentum give rise to large QCD corrections.  In order to suppress
these enhanced corrections
\cite{Rubin:2010xp} we apply an additional cut on the final state in our
analysis. Focussing on the leptonic final state we have to make sure
that the gauge-boson pairs or its decay products, the four leptons, are
not produced with too much transverse momentum from the recoil
against a system of strongly interacting particles.  We achieve this by 
applying a condition on the leptons' transverse  momenta $\vec \ell_{i,\rT}$ in the
laboratory frame, 
\begin{equation}
  \label{eq:cut}
  \left|\sum_i \vec \ell_{i,\rT}+ \vec{p}_{\rT,\mathrm{miss}} \right| < \rho \left ( \sum_i \left|
    \vec\ell_{i,\rT} \right| + |\vec{p}_{\rT,\mathrm{miss}}| \right)\,.
\end{equation}
Here, we consider all visible leptons $i$ from EW boson decays, i.e.\
$i$ runs to 2, 3 or 4 in case of WW, WZ or $\mathrm{ZZ}$ events. We
additionally assume that the missing transverse momentum in the event
solely stems from neutrinos from W decays. The left-hand-side
of Eq.~(\ref{eq:cut}) is small whenever the system has a small
transverse momentum, as the leptons from the EW system recoil
against each other.  After some experimenting, we find that $\rho = 0.3$
gives a good selection of events while leaving enough events for our
analysis at the same time.  

\begin{figure}
  \centering
  \includegraphics[width=0.49\textwidth]{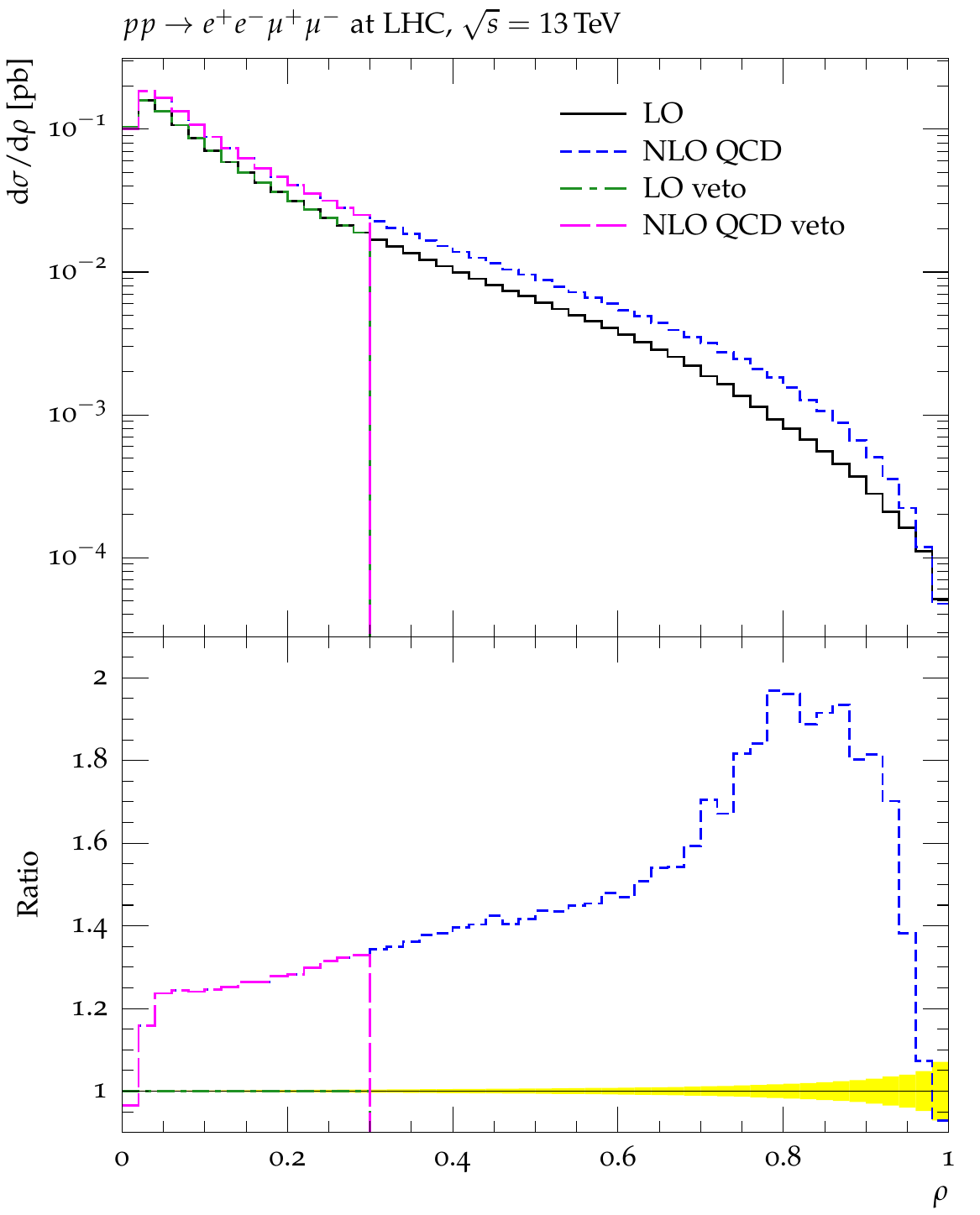}
  \includegraphics[width=0.49\textwidth]{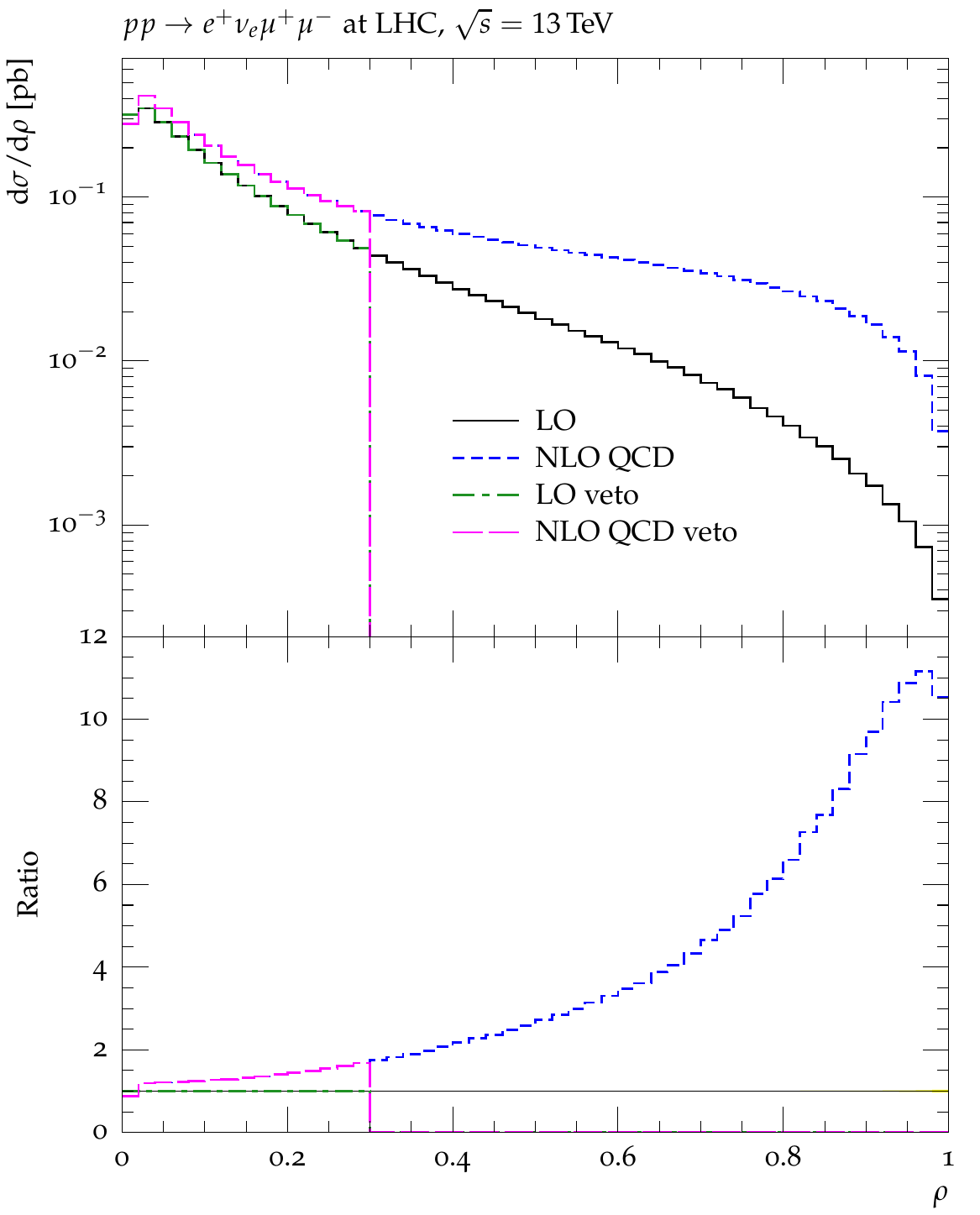}
  \\[2em]
  \includegraphics[width=0.49\textwidth]{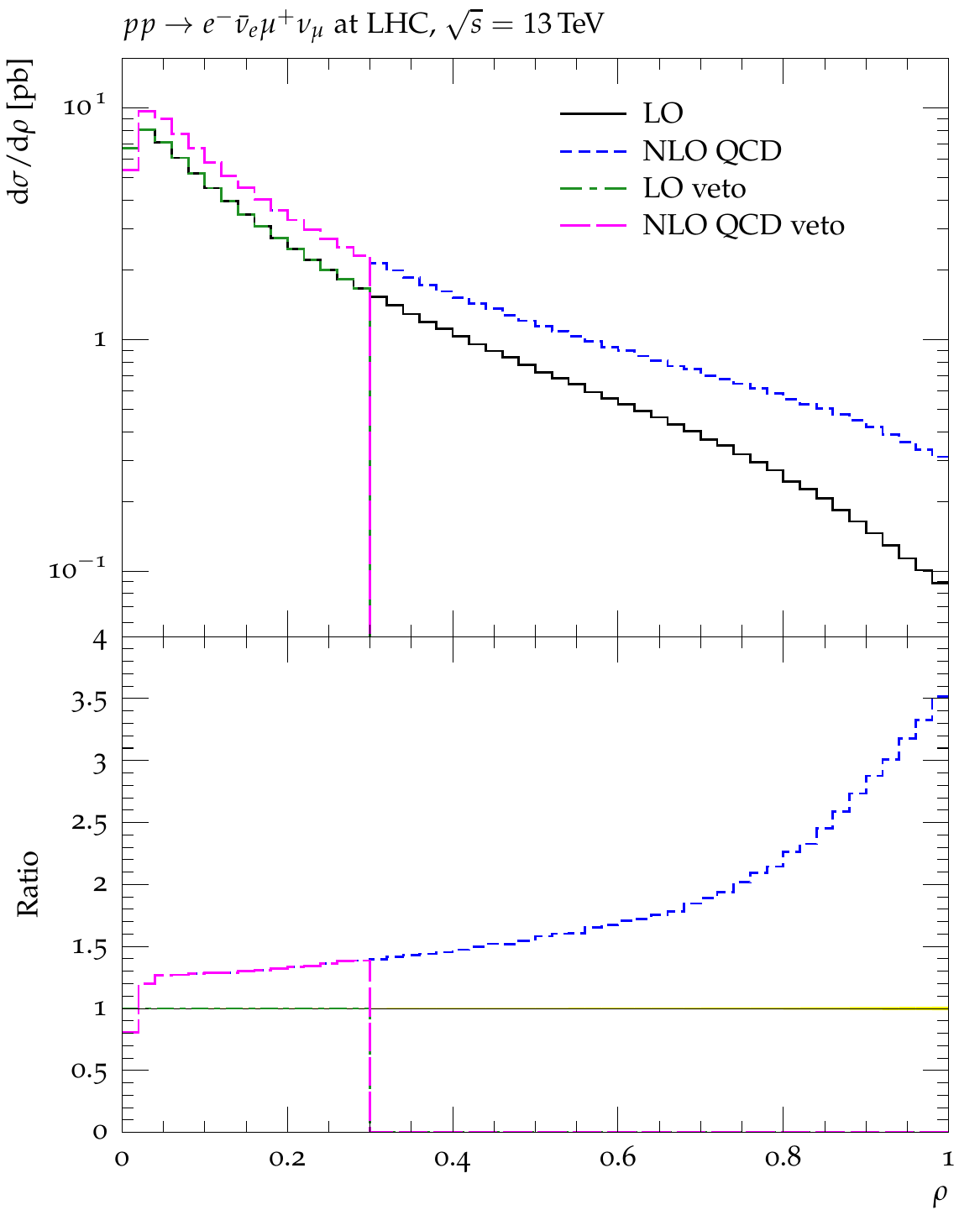}
  \caption{Histograms in the veto variable $\rho = \left|\sum_i \vec
      \ell_{i,\rT}+ \vec{p}_{\rT,\mathrm{miss}} \right|  /  \left ( \sum_i \left|
    \vec\ell_{i,\rT} \right| + |\vec{p}_{\rT,\mathrm{miss}}| \right)\,$ for the three
    different final states under consideration.\label{fig:veto} For
    $\rho<0.3$ LO and LO veto coincide, as well as NLO QCD and NLO QCD veto.}
\end{figure}
In Fig.~\ref{fig:veto} we show histograms of the ratio $\rho$ of vector
and scalar sums of lepton transverse momenta that we finally apply the
cut on.  We show runs with and without NLO QCD corrections. We find that
the chosen value $\rho=0.3$ is a sensible choice for all three
combinations of boson pairs.  A good number of events is still available
for the analysis while at the same time we find that the events with
large distortions from hard radiation are vetoed with our selection.
These are peaked at large values of the ratio in all three cases.
In a full experimental analysis the value of $\rho$ might be subject to
optimization for the individual cases of vector boson pairs.

\begin{figure}[p]
  \centering
  \includegraphics[width=0.49\textwidth]{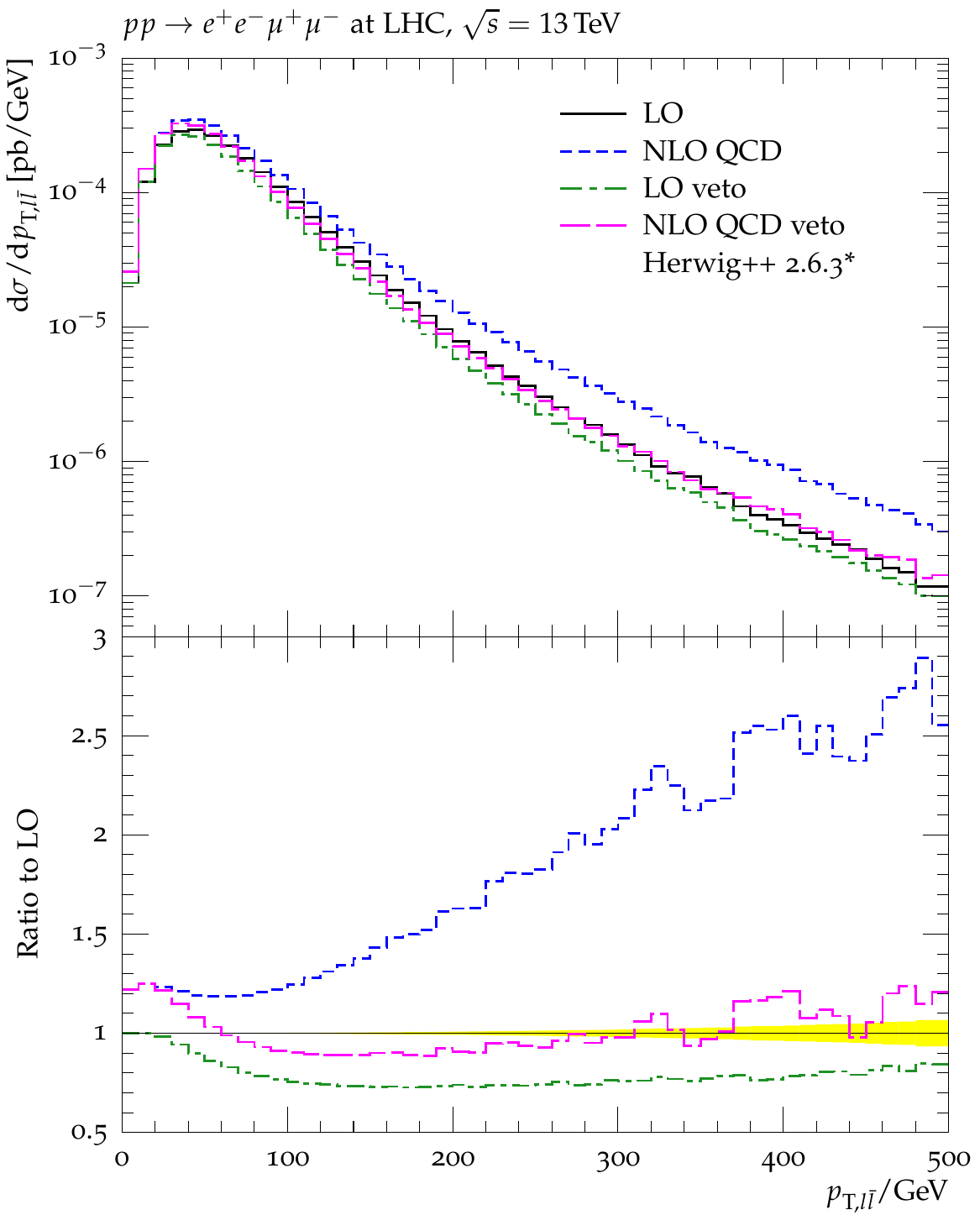}
  \includegraphics[width=0.49\textwidth]{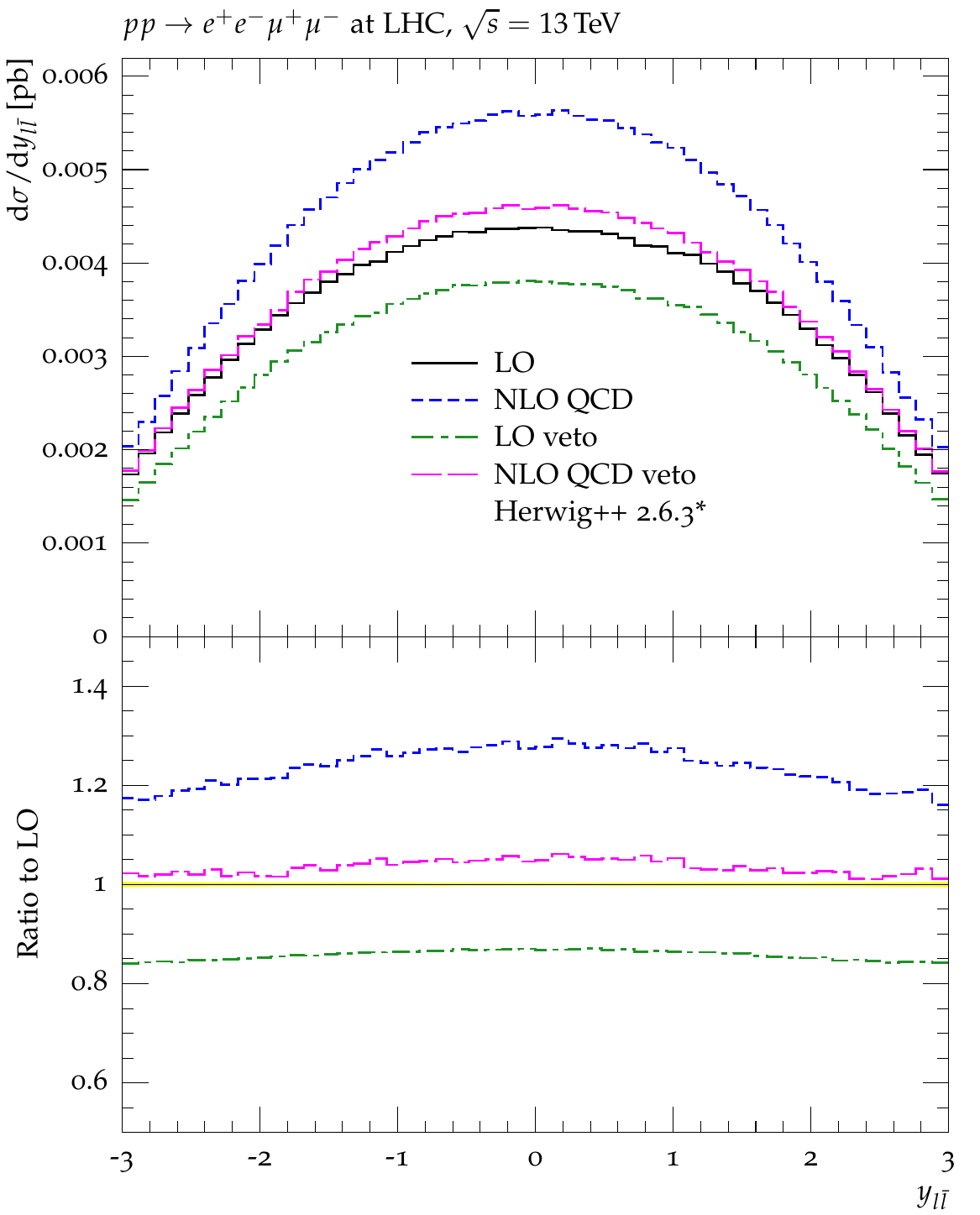}
  \caption{ZZ production at LHC13, effect of
    veto on selected leptonic observables. \label{fig:zz8veto}}
\end{figure}

\begin{figure}[p]
  \centering
  \includegraphics[width=0.49\textwidth]{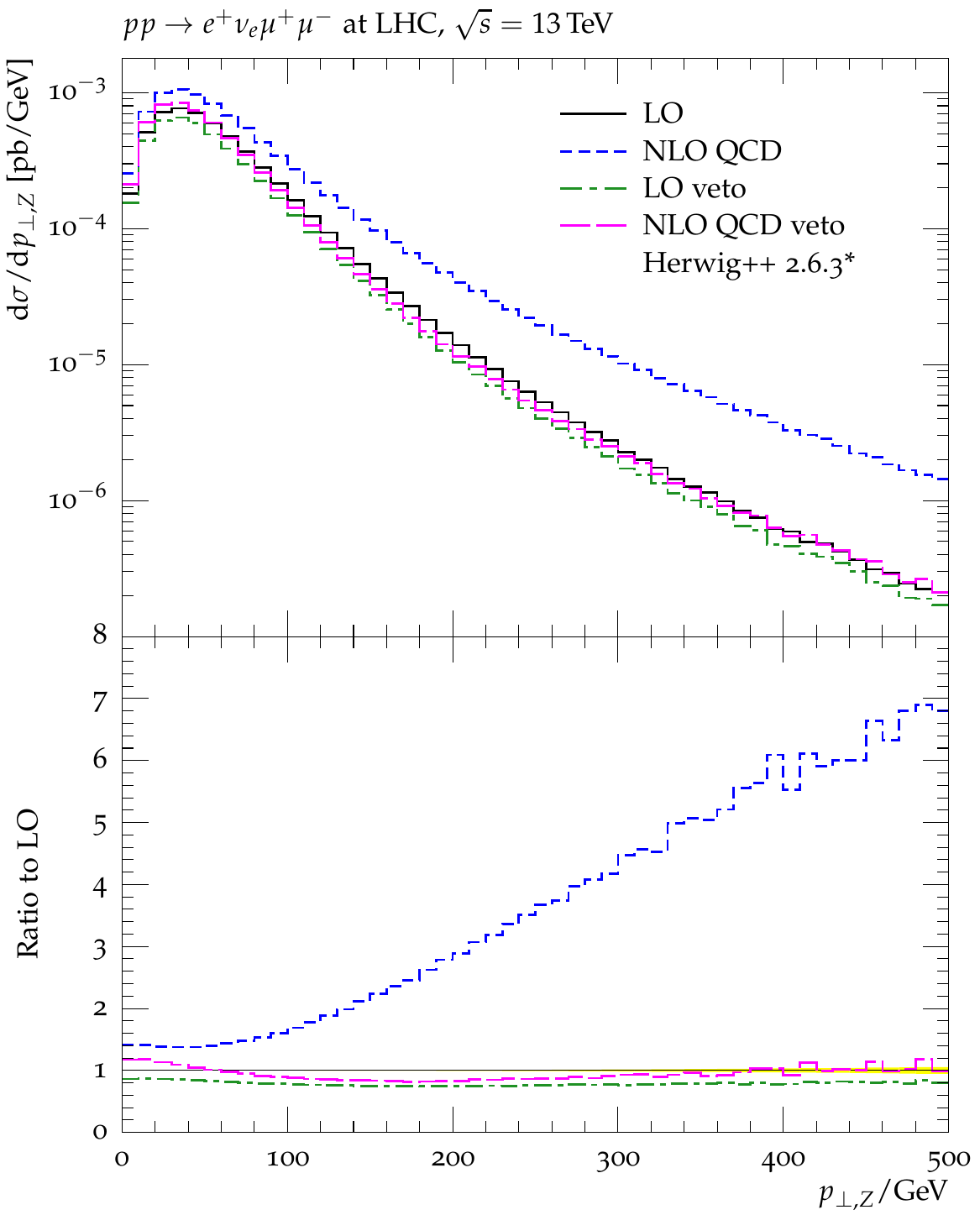}
  \includegraphics[width=0.49\textwidth]{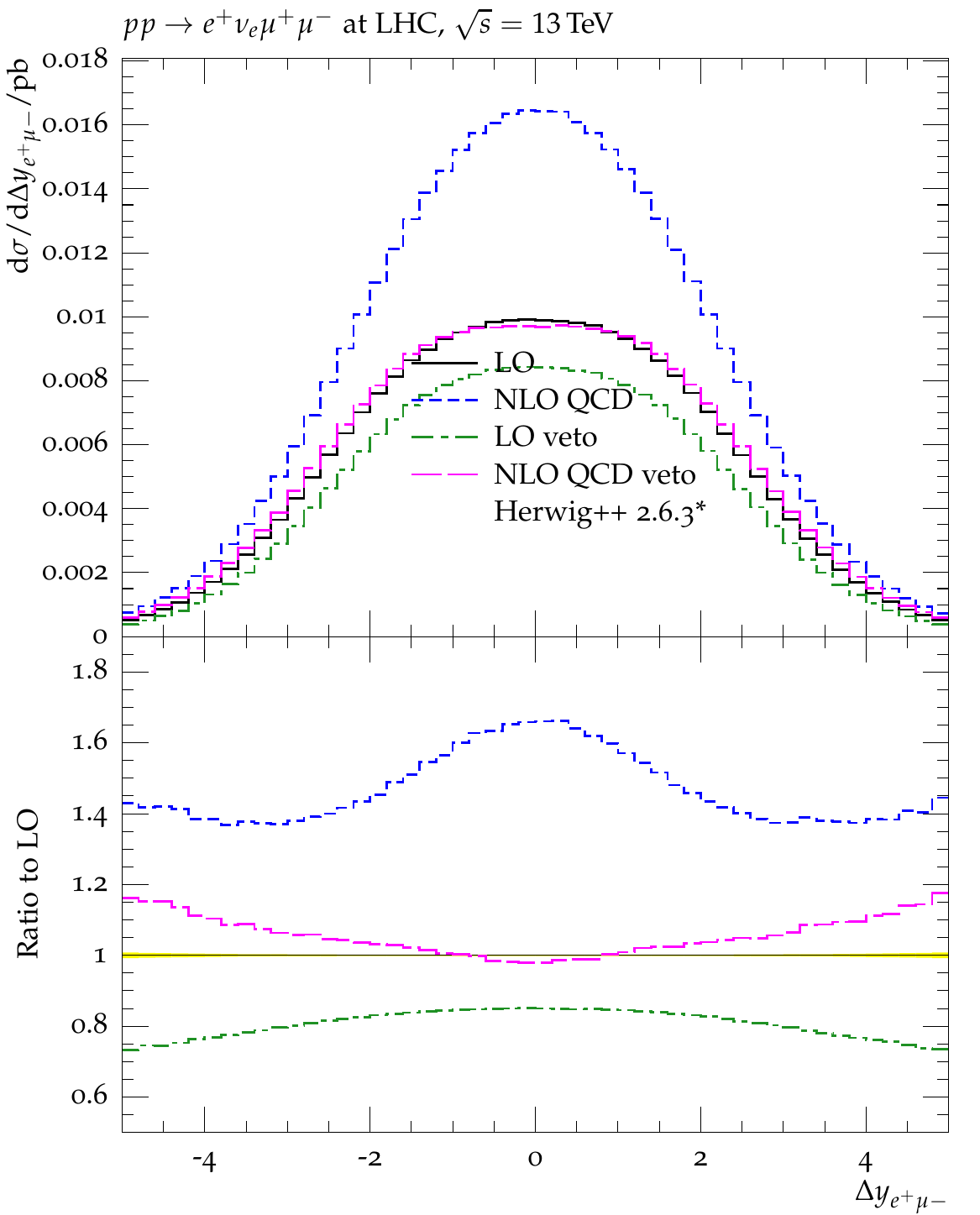}
  \caption{WZ production at LHC13, effect of
    veto on selected leptonic observables. \label{fig:wz8veto}}
\end{figure}

\begin{figure}[p]
  \centering
  \includegraphics[width=0.49\textwidth]{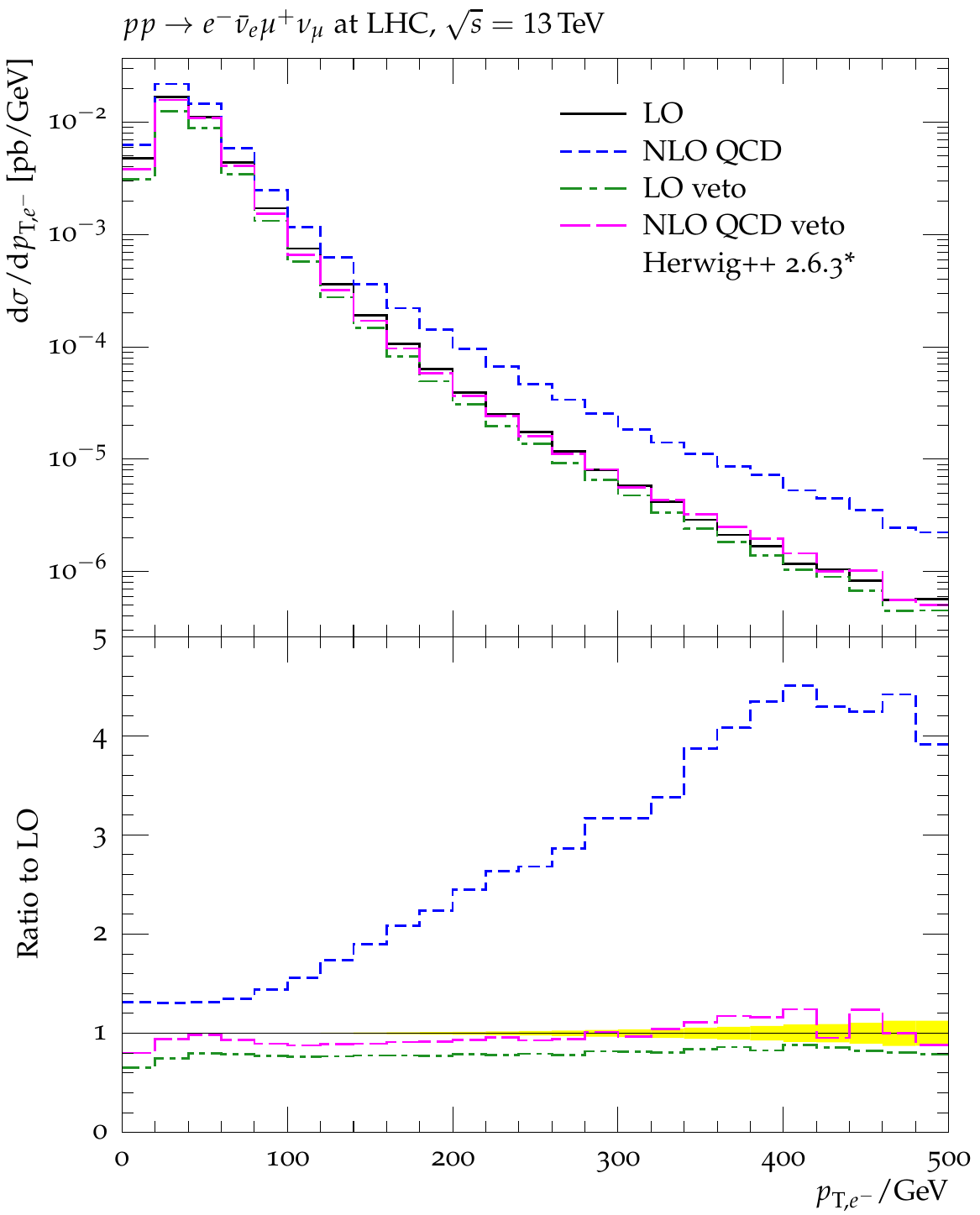}
  \includegraphics[width=0.49\textwidth]{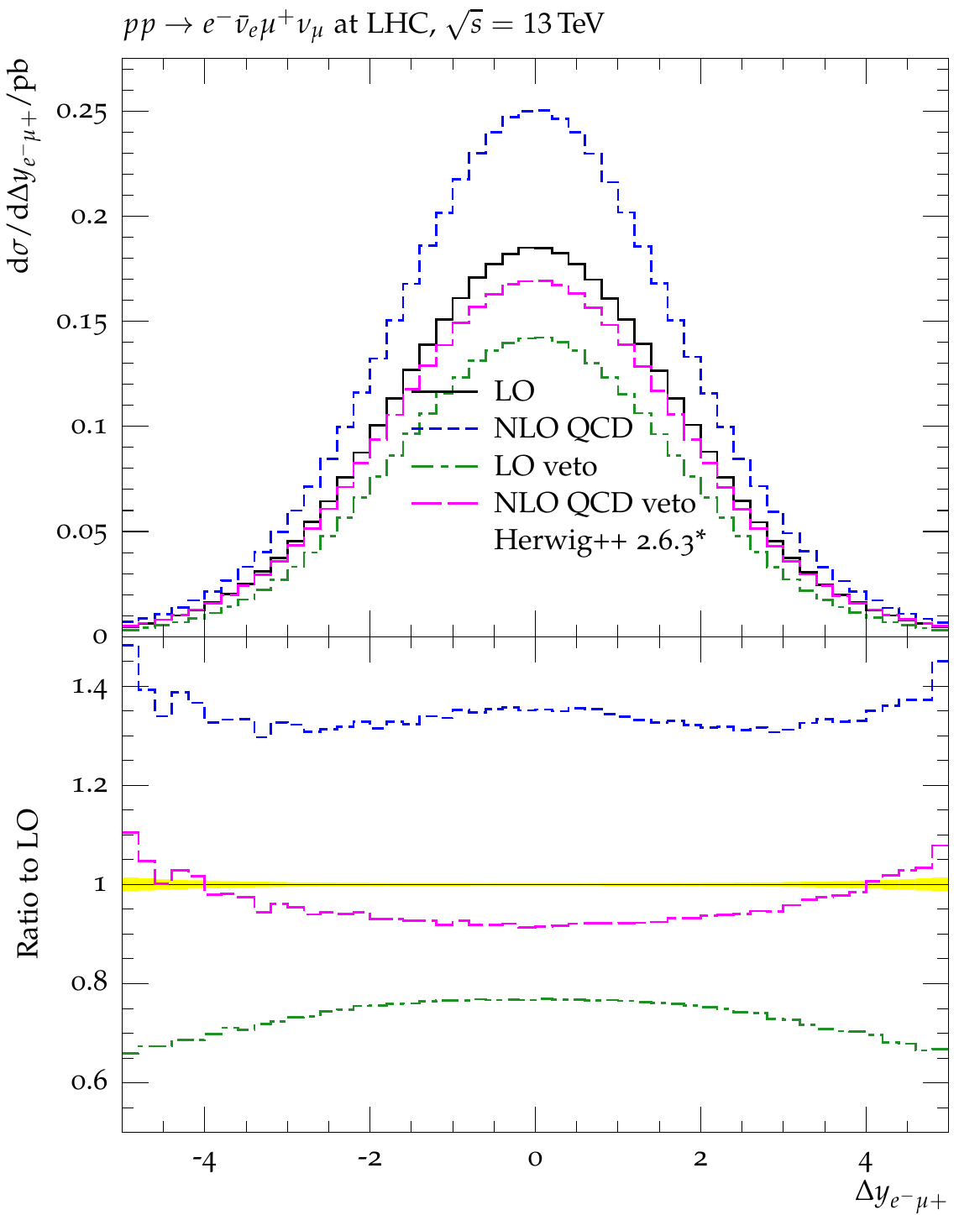}
  \caption{WW production at LHC13, effect of
    veto on selected leptonic observables.
    \label{fig:ww8veto}}
\end{figure}

In addition we show actual leptonic observables with and without
application of the lepton veto in
Figs.~\ref{fig:zz8veto}-\ref{fig:ww8veto}.  
Let us compare LO and NLO predictions for the $p_{\rT,l\bar{l}}$
distribution in ZZ production (Fig.~\ref{fig:zz8veto}, left) which
corresponds essentially to the transverse-momentum distribution of the
Z-boson. Without cut the NLO distribution exceeds the LO distribution by
a factor of two at $p_{\rT,l\bar{l}} = 300$ GeV and more at larger
transverse momenta. Introducing the cut~\eqref{eq:cut} with $\rho=0.3$
removes most of this excess such that the difference between LO and NLO
distributions is reduced to $\mathcal{O}(20\%)$ and remains relatively
constant as function of $p_{\rT,l\bar{l}}$. The ratio between the NLO
and LO rapidity distribution, in contrast, is fairly constant with a
ratio of NLO/LO ${}\sim 1.2$, and is reduced to ${}\sim 1$ by the cut
for $\rho=0.3$. A similar behaviour is observed for WZ
(Fig.~\ref{fig:wz8veto}) in the $e^+\nu_e\mu^+\mu^-$ mode with a giant
correction factor of 4 for transverse momenta of the Z boson of 300 GeV
and a correction between 1.4 and 1.6 for the rapidity distribution. As
before, one obtains a significant reduction of the QCD correction down
to $\mathcal{O}(20\%)$ by introducing the cut~\eqref{eq:cut}, and a
similar behaviour is observed for W$^+$W$^-$ production in the
$e^-\bar{\nu}_e\mu^+\nu_\mu$ channel (Fig.~\ref{fig:ww8veto}). In total,
after applying the cut, LO and NLO results are close over the whole
range in transverse momentum and rapidity.

In order to assess the sensitivity to the choice of partonic kinematic
variables we have studied the impact of $K(\hat s, \hat t)$ with two additional
choices for the kinematical variables for ZZ  production, as in this
case we have full access to the kinematics of the final state.

Let us call the variables from the direct access to the kinematics
within \HERWIG{} before the parton shower $(\hat s_{\rm hard}, \hat
t_{\rm hard})$. As an alternative, we reconstruct the kinematics from
the vector-boson final state as outlined in the previous section. Here,
no knowledge of the initial-state partons is needed for the computation
of the kinematic invariants. $\hat s_{\rm rec}$ is given by the invariant
mass of the vector-boson pair and $\hat t_{\rm rec}$ is reconstructed
from the scattering angle in the centre of mass frame as proposed in
Eqs.~\eqref{eq:m4l}--\eqref{eq:costheta}.  For illustration, we also
consider a choice of variables that is very likely to be wrong.  We take
the initial state partons after the termination of the parton shower.
This parton pair will have a much larger invariant mass $\hat s_{\rm
  PS}$ due to the parton showering.  We then find the four momenta of
the outgoing vector bosons and compute $\hat t_{\rm PS}$ with respect to
this initial state.  This last choice will demonstrate us the
sensitivity to the parton shower emissions.  For each event produced by
\HERWIG{} we construct three sets of kinematic variables $(\hat s_i,
\hat t_i)$ with $i\in \{{\rm hard}, {\rm rec}, {\rm PS}\}$.  Then we
compute the three different $K$ factors $K_i$ from these three sets of
variables.  Taking $K_{\rm hard}$ as a reference we compute the ratio
$K_i/K_{\rm hard}$ for each event.  If the reconstruction would be
insensitive to the choice of kinematics this ratio ought to be unity all
the time.

\begin{figure}[t!]
  \centering
  \includegraphics[width=0.7\textwidth]{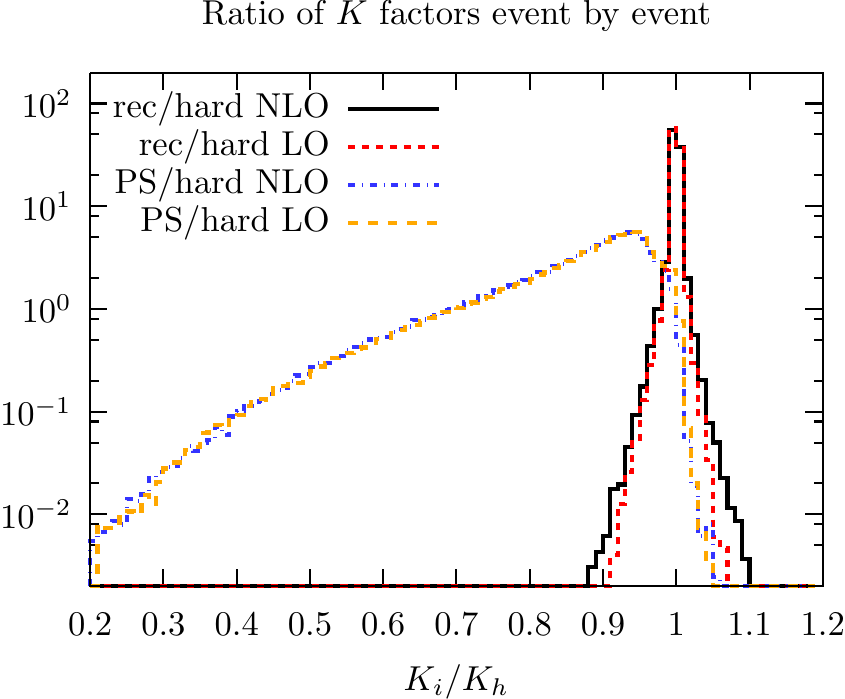}
  \caption{The ratio of $K$ factors for different reconstructions $i$ of
    the kinematic variables $(\hat s_i, \hat t_i)$ in Z-pair production
    at LHC13.
    See text for further explanation.}
  \label{fig:reconstruction}
\end{figure}

In Fig.~\ref{fig:reconstruction} we show a histogram of this ratio for
the two different ratios $K_{\rm rec}/K_{\rm hard}$ and $K_{\rm
  PS}/K_{\rm hard}$.  As expected, the results from the reconstruction
after the application of the parton shower are quite different from
unity and show a very broad peak.  This is reasonable and shows us that
a wrong choice of kinematic variables gives a sensitive difference in
the computation of the $K$ factor.

In strong contrast, the EW $K$ factors after the kinematics
reconstruction from the final state compared to the $K$ factor from the
variables of the hard process inside the event generator are very close
and their ratio shows a very strong peak around unity, showing us that
the two prescriptions lead to nearly identical results. 

In the same figure~\ref{fig:reconstruction} we compare the results from
runs with and without NLO QCD corrections applied.  In both cases we get
very similar results for the EW $K$ factor.  The additional hard gluon
from the real emission graph distorts the kinematics that was
reconstructed in the leading order case only slightly.  As expected the
QCD corrections give a slightly bigger difference between the two
reconstruction schemes.

We conclude that our computation of the EW correction is very robust
against small variations of the kinematics as long as the variables are
sensibly chosen and a sensible veto is applied.  The reconstruction of
kinematics from the final state alone, as proposed in the previous
section, is a viable choice.  One should, however, be careful, as e.g.\
naively ignoring the parton shower in the reconstruction of kinematics
will lead to significantly different results.

\section{\HERWIG~results}\label{se:herwigresults}

Having established the role of the veto 
on the leptonic final state 
 we  present results of differential
distributions in leptonic observables.  We apply the same selection of
final states as described above in Sect.~\ref{se:4ldetails} and
Sect.~\ref{se:herwig}.  Now, the leptons are, however, selected from a
full hadronic final state with an additional isolation cut of $R=0.2$.
Furthermore, the final state is of course modified by parton showers and
hadronization as well as additional soft and collinear photon radiation.
We left the underlying event switched off as we expect only a small
effect for the observables presented here.  In all cases we have
generated $10$M unvetoed events.  The leptonic veto has only been
applied at the analysis level.

\begin{figure}[t]
  \centering
  \includegraphics[width=0.49\textwidth]{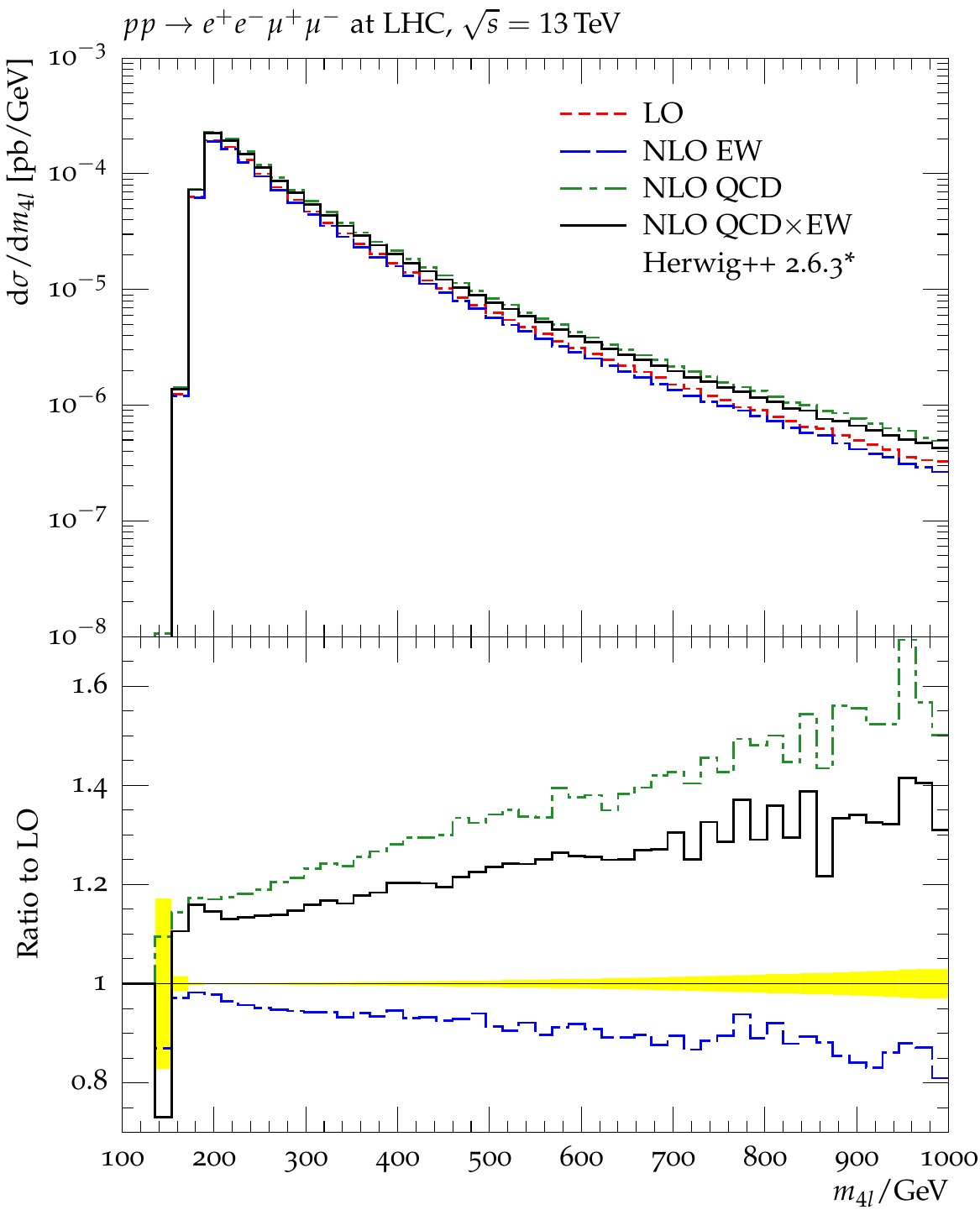}
  \includegraphics[width=0.49\textwidth]{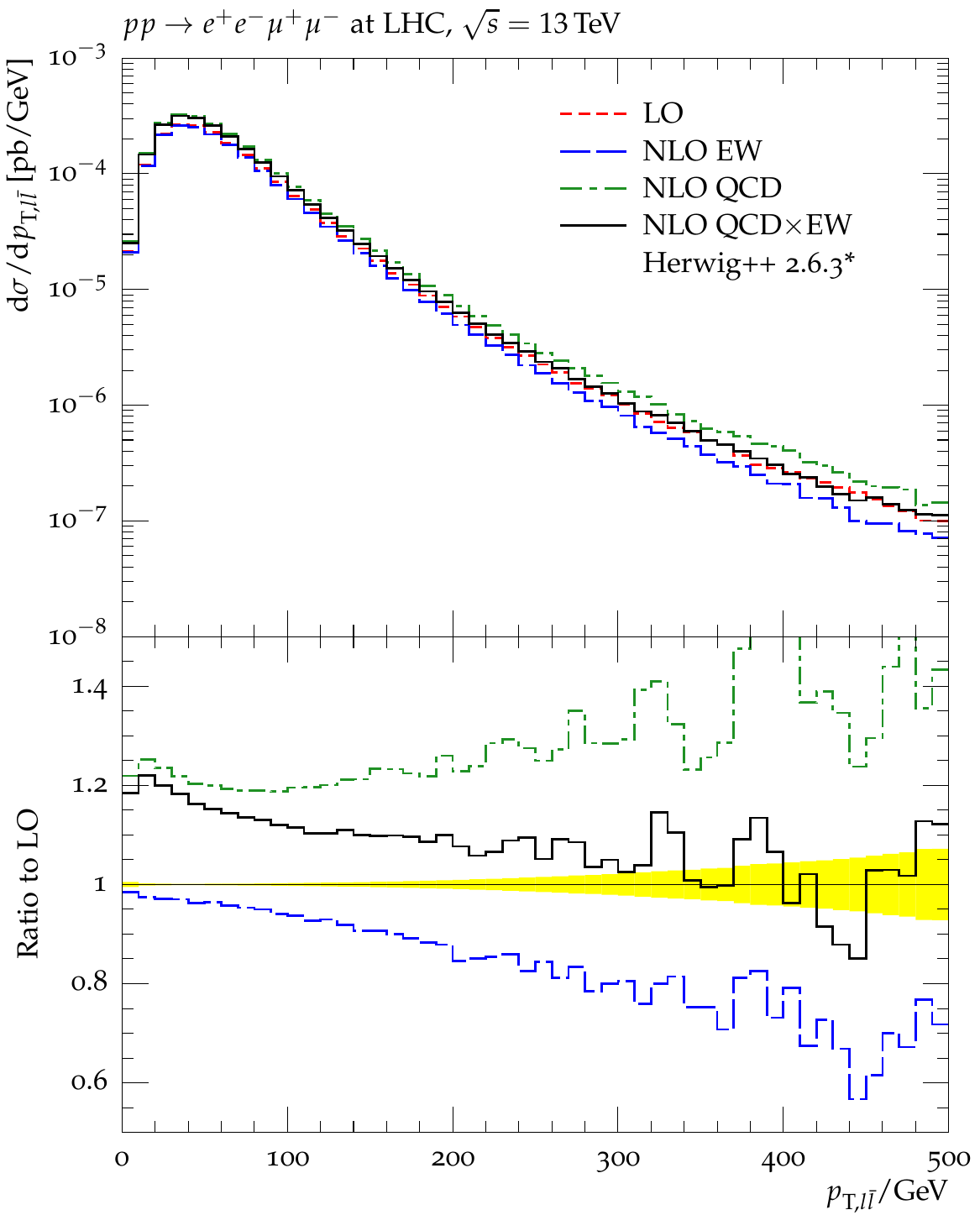}
  \\[2em]
  \includegraphics[width=0.49\textwidth]{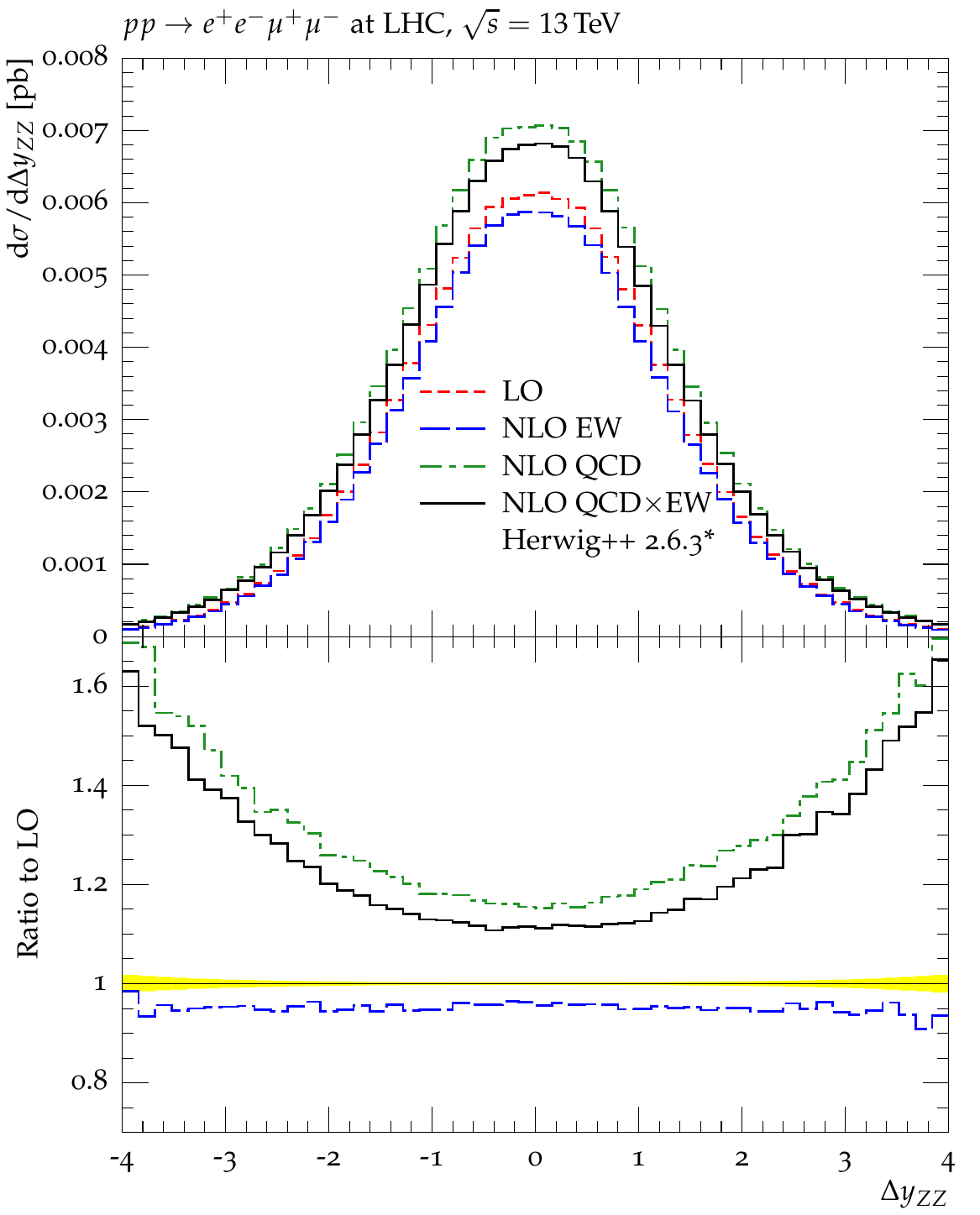}
  \includegraphics[width=0.49\textwidth]{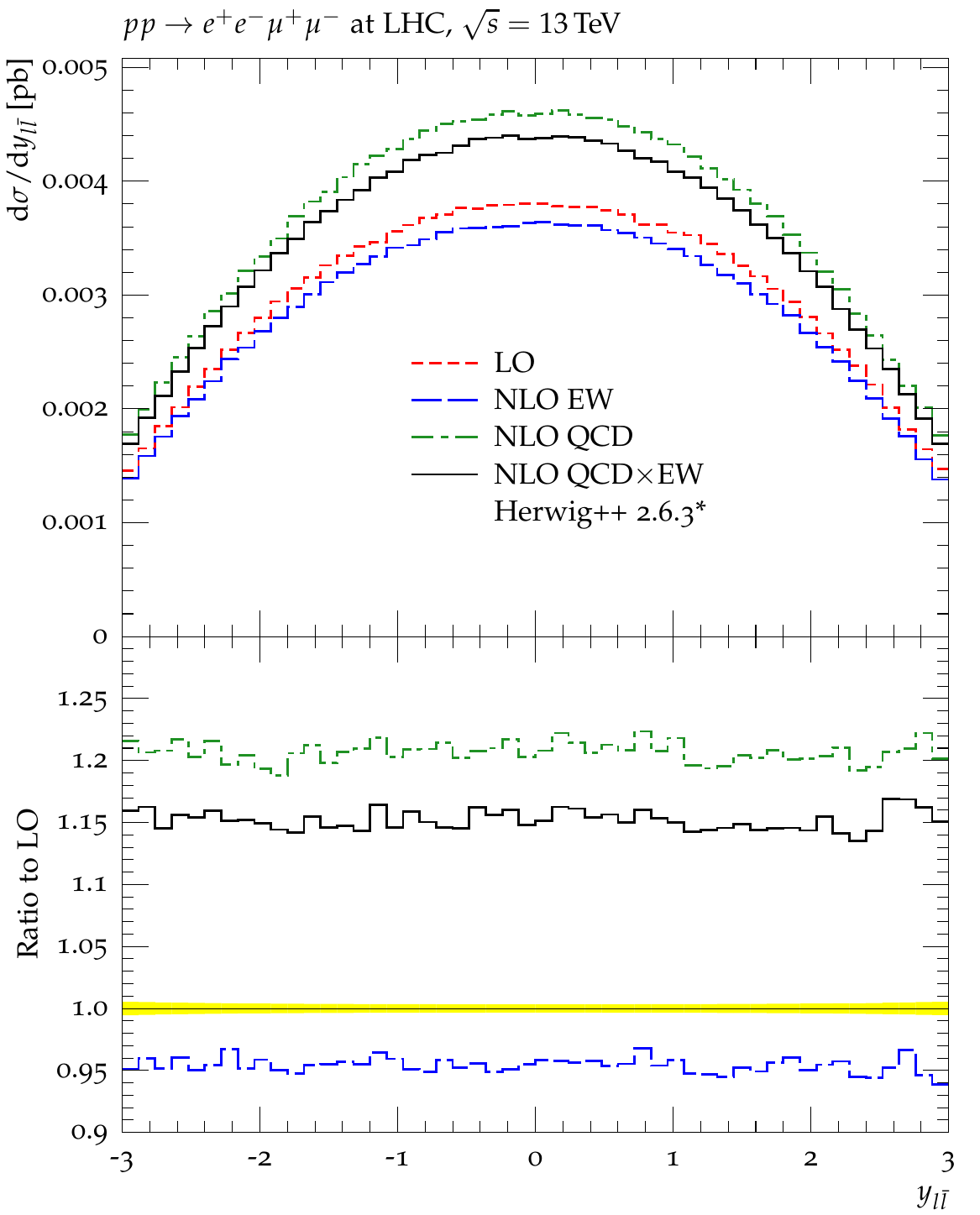}
  \caption{Results from ZZ events at $\sqrt{s}=13$\,TeV. We show
    observables that can be reconstructed from the leptonic final state
    after we applied the veto Eq.~(\ref{eq:cut}).
    \label{fig:zz8ew}}
\end{figure}

In Fig.~\ref{fig:zz8ew} we show a number of observables for the final
state of $\mathrm{ZZ}$ production.  In all cases we show four lines:
the leading order result (LO), results with only electroweak (NLO EW)
or QCD corrections (NLO QCD) applied and, finally, the result with
the combined EW and QCD corrections in the multiplicative scheme as
outlined above.  In both cases, with and without NLO QCD corrections,
the additional EW correction is as sizable as in the partonic case,
i.e.\ of the order of $-5\,\%$ for small $m_{4l}$, reaching up to
$-20\,\%$ for $m_{4l}$ close to 1 TeV or $p_{\rT,l\bar{l}}$ close to 500
GeV. The rapidity distributions, shown in the lower two plots of
Fig.~\ref{fig:zz8ew}, receive the correction of $-4\%$ typical for the
low-$\hat{s}$ configuration. 

\begin{figure}[t]
  \centering
  \includegraphics[width=0.49\textwidth]{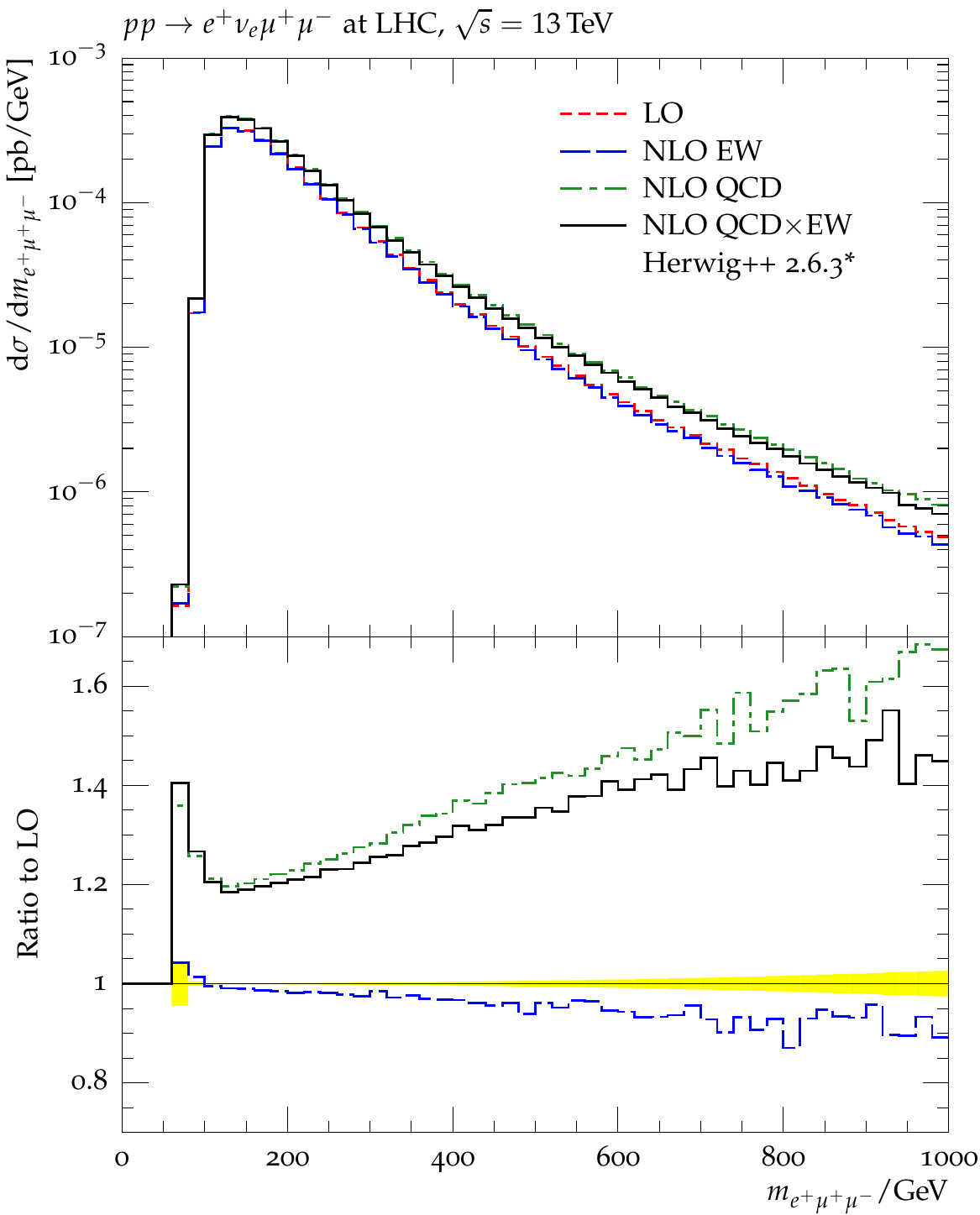}
  \includegraphics[width=0.49\textwidth]{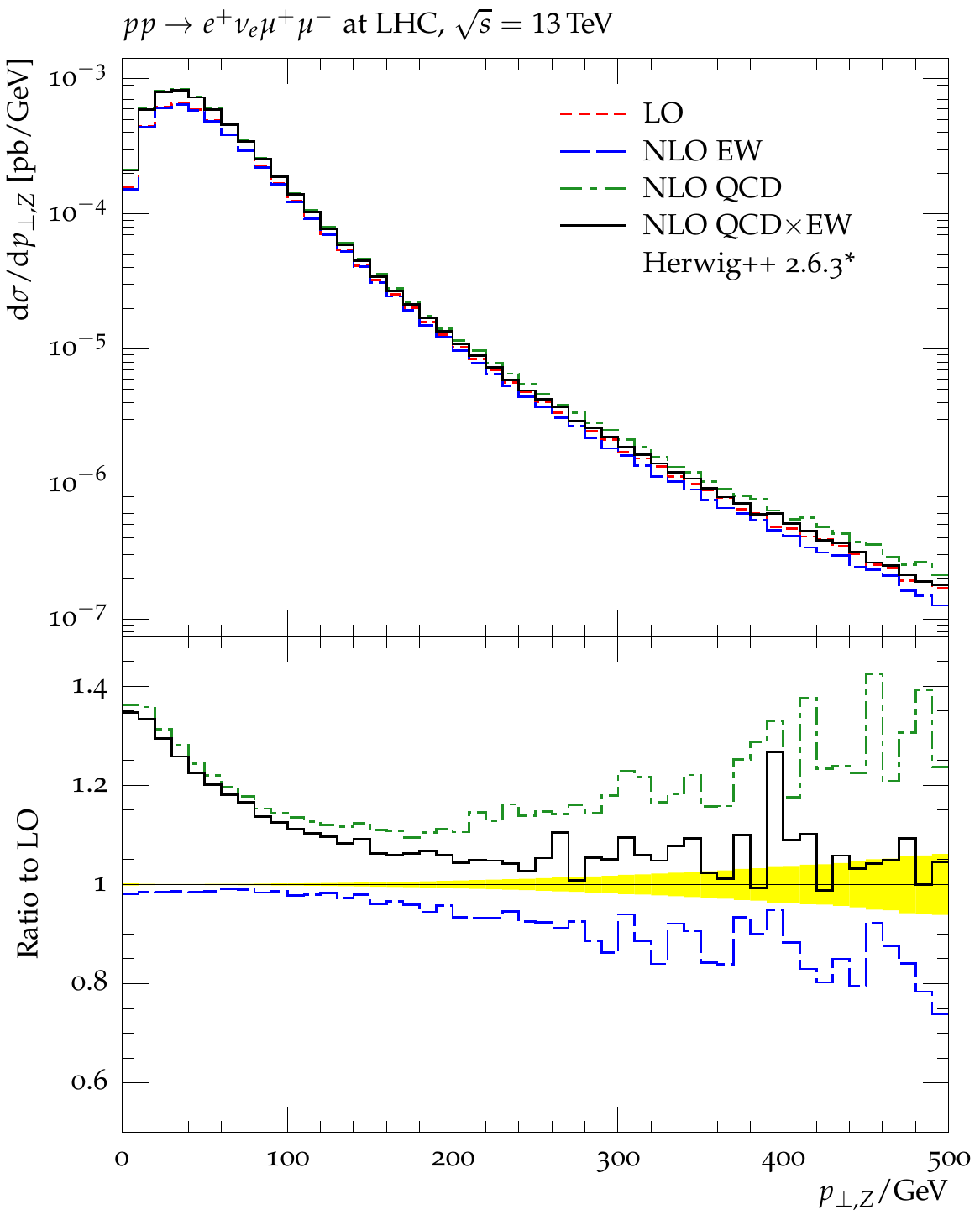}
  \\[2em]
  \includegraphics[width=0.49\textwidth]{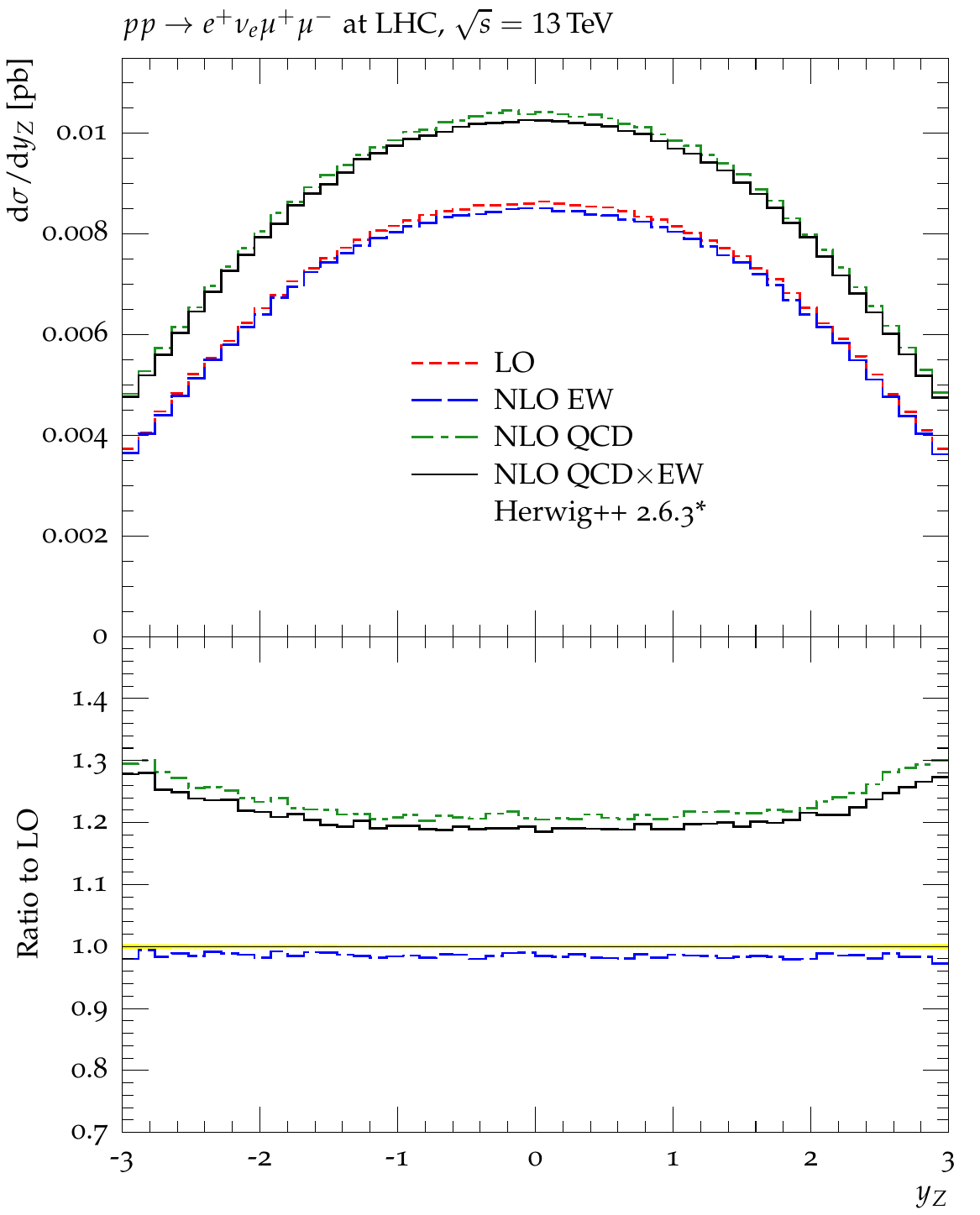}
  \includegraphics[width=0.49\textwidth]{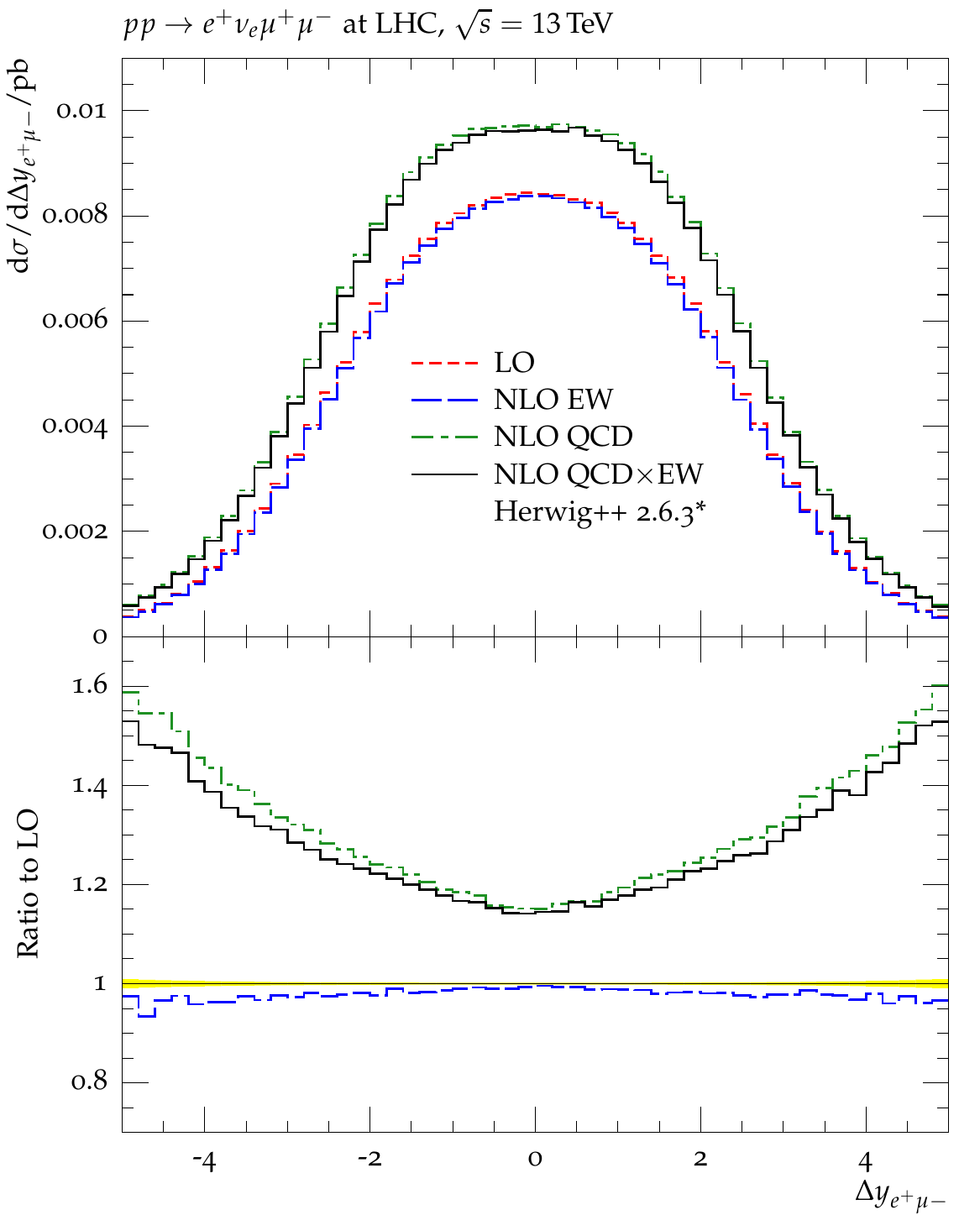}
  \caption{Results from WZ events at $\sqrt{s}=13$\,TeV. We show
    observables that can be reconstructed from the visible 
    leptonic final state
    after we applied the veto Eq.~(\ref{eq:cut}).
    \label{fig:wz13ew}}
\end{figure}

A similar picture emerges in the case of WZ production which we show in
Fig.~\ref{fig:wz13ew} for $\sqrt{s}=13\,$TeV.  Here, the EW corrections
are smaller than in the ZZ case.  In every observable we find that the
EW corrections act quite similar on the final states with and without
NLO QCD corrections.  While the QCD corrections are moderate, at the
level of 20\,\%, the EW corrections vary between zero and $-20\,\%$, and
are again sizable for high transverse momenta of the Z bosons.

\begin{figure}[t]
  \centering
  \includegraphics[width=0.49\textwidth]{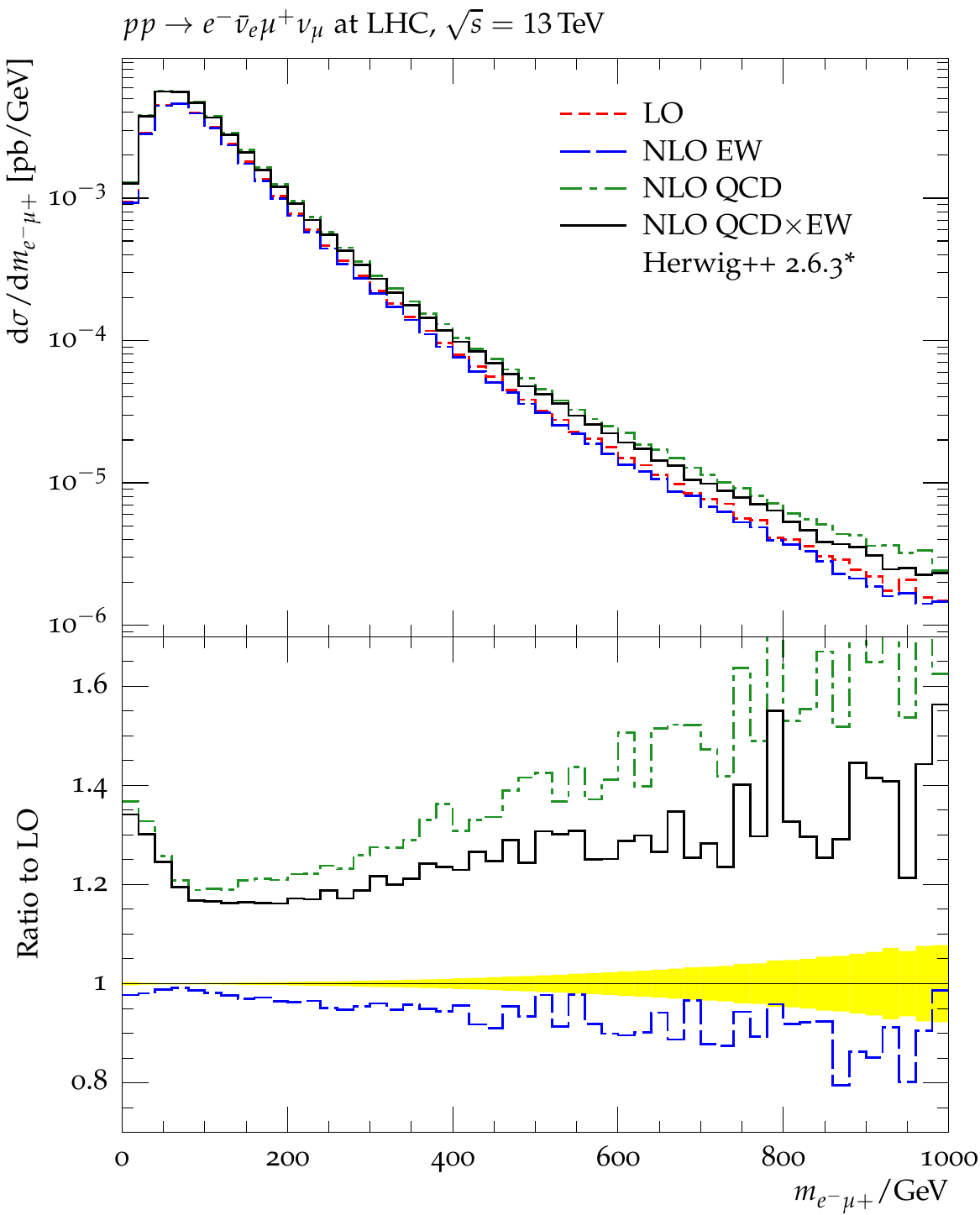}
  \includegraphics[width=0.49\textwidth]{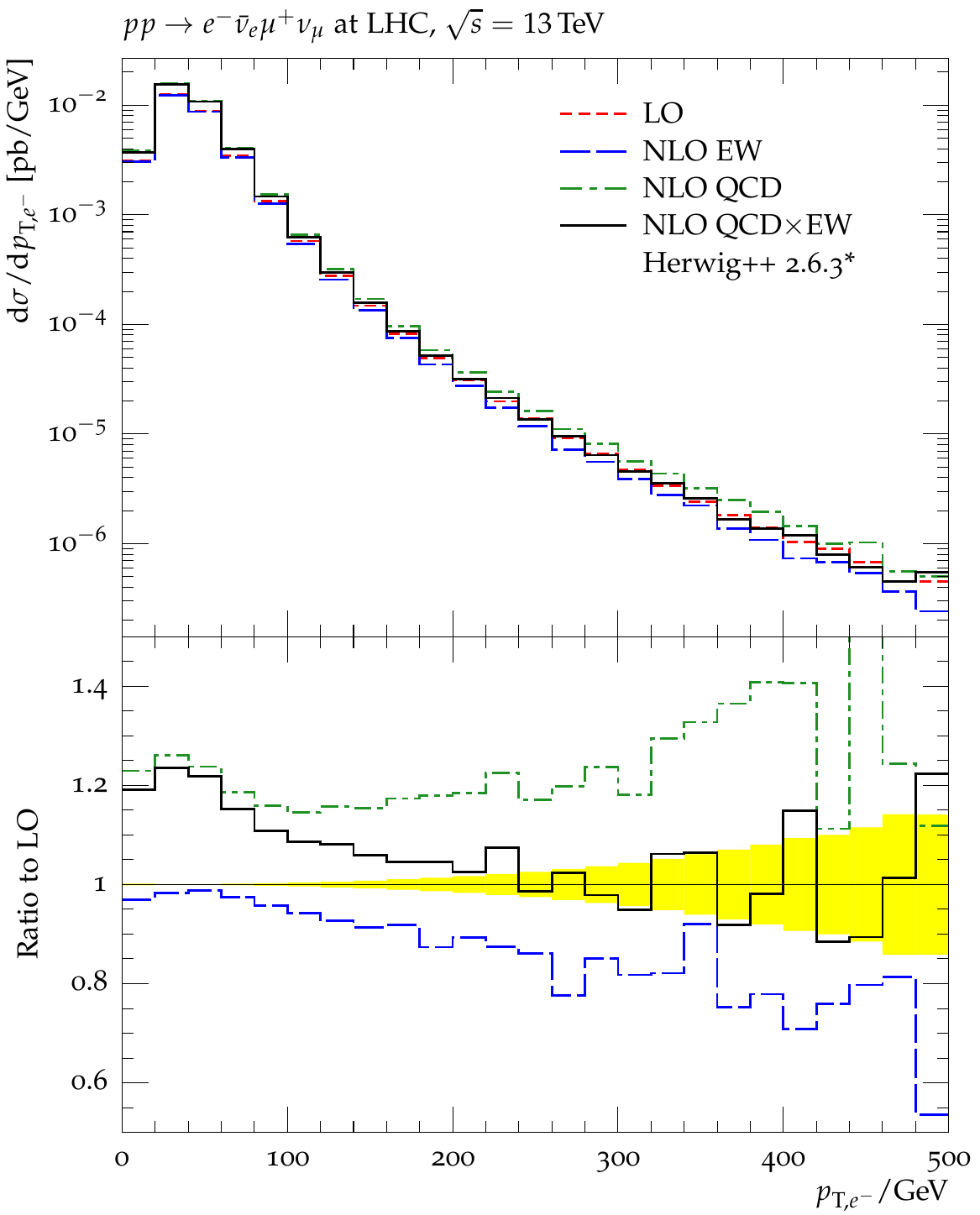}
  \\[2em]
  \includegraphics[width=0.49\textwidth]{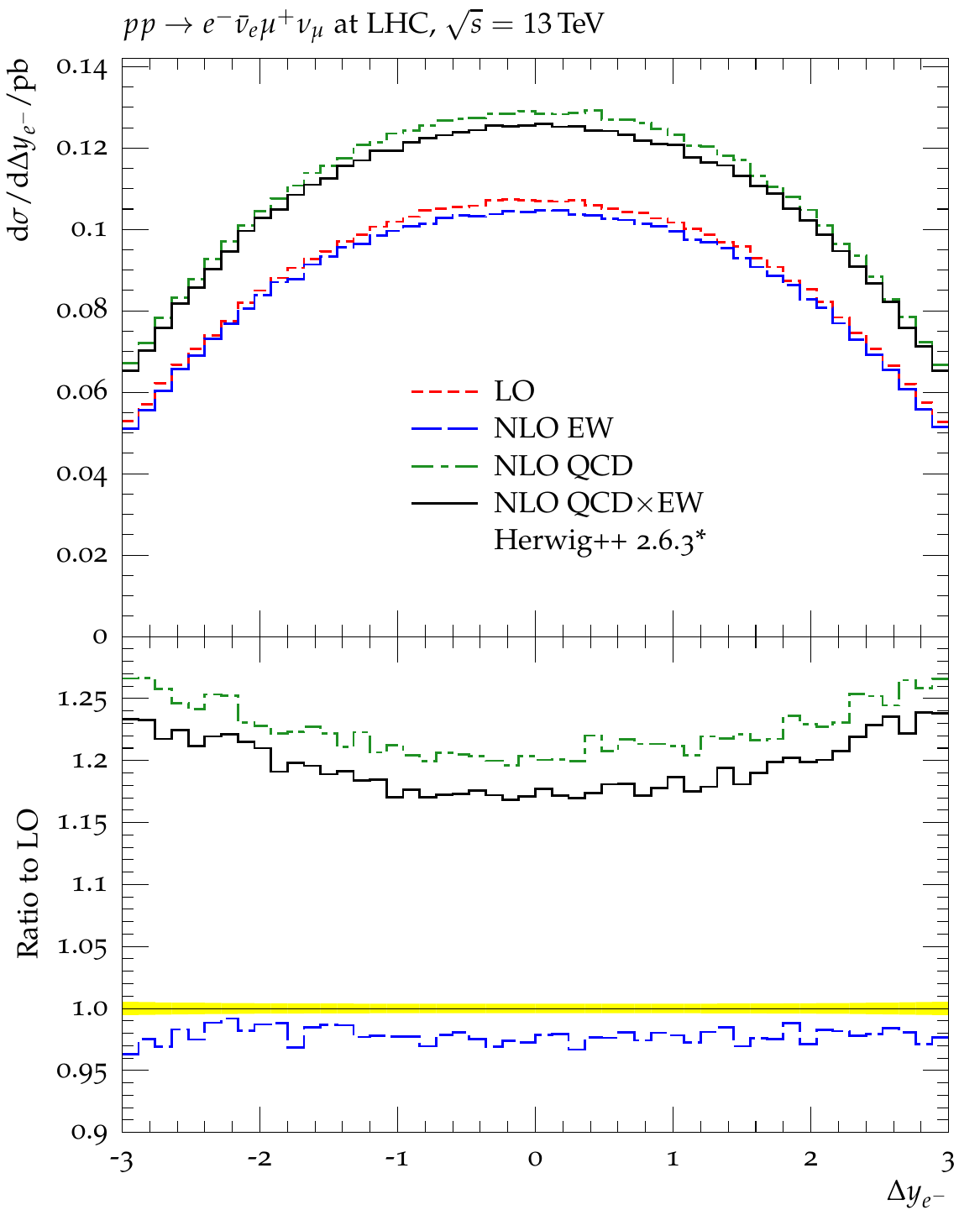}
  \includegraphics[width=0.49\textwidth]{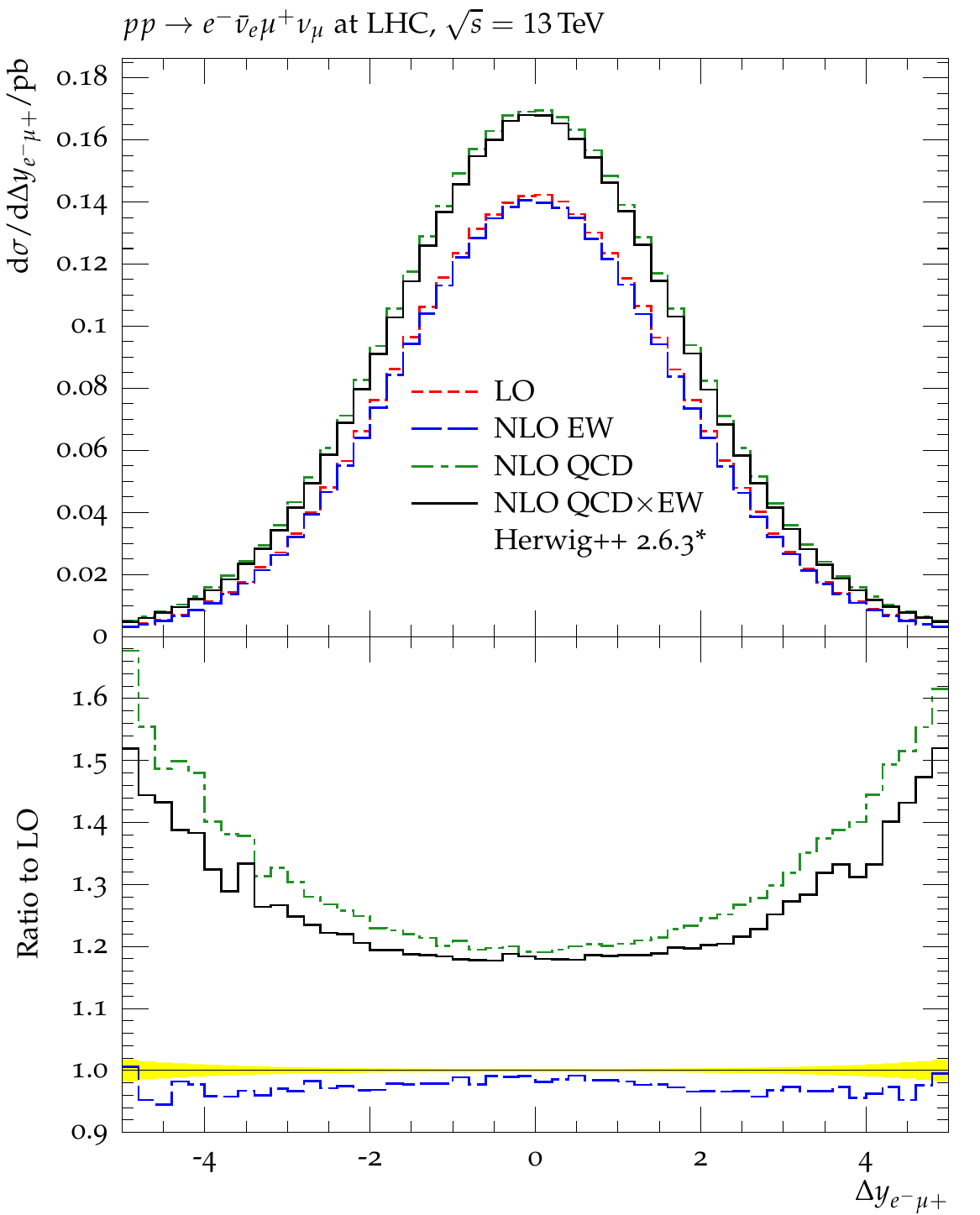}
  \caption{Results from WW events at $\sqrt{s}=13$\,TeV. We show
    observables that can be reconstructed from the visible leptonic
    final state after we applied the veto
    Eq.~(\ref{eq:cut}). \label{fig:ww8ew}}
\end{figure}

Finally, in Fig.~\ref{fig:ww8ew} we consider selected observables for the
case of W-pair production at 13\,TeV.  Here, the EW corrections are slightly
larger than in the WZ case but the overall picture remains the same.
The QCD corrections are quite large but tamed by our veto on the
leptonic final state, typically of the order of 20\,\%.  The EW
corrections are typically of the order of 5\,\% but again sizable in the
case of large lepton transverse momentum, where they can
completely compensate the QCD corrections.

\section{Summary and Conclusions}
We have computed the full NLO EW corrections to resonant vector-boson
pair production at the LHC, taking into account leptonic decays and
corresponding spin correlations. We propose a simple and
straight-forward method -- relying on unpolarized $2\to 2$ $K$-factors
-- to implement our results in any state-of-the-art MC generator. The EW
corrections are combined with precise QCD predictions in a
multiplicative approach. We emphasize that our method also allows for an
\emph{a posteriori} implementation of EW $K$-factors into MC samples
that have already been generated. To demonstrate the practicability of
our approach, we have included our corrections in the \HERWIG{} MC
generator and presented various distributions for four-lepton production
at the LHC obtained in the \HERWIG{} setup, including EW corrections,
NLO QCD corrections matched to parton showers, as well as hadronization
effects. In the future our method could also be applied to other process
classes, such as $V$+jet production at the LHC, allowing for
phenomenological studies that combine EW precision on the one hand and an
adequate treatment of dominating QCD effects on the other.

\begin{appendix}
\section{Endpoint contributions from dipoles}
According to Ref.~\cite{Dittmaier:1999mb} the endpoint contribution
compensating soft and collinear QED singularities from the virtual
corrections of the unpolarized $V$-boson pair production process $q\bar{q}' \to V_1
V_2$ is given by the general expression
\begin{equation}\label{eq:endpt}
  \rd \sigma_{q\bar{q}' \to V_1V_2\gamma}^{\mathrm{E}} =
  -\frac{\alpha}{2\pi}\sum_{I\neq J} Q_I \sigma_I Q_J \sigma_J
  G_{IJ}^{(\mathrm{sub})}(P_{IJ},m_I,m_J) \; \rd \sigma^{\LO}_{q\bar{q}'\to V_1V_2}(p_I,p_J)\,,
\end{equation}
where the sum runs over all dipole contributions $(IJ)$ of charged
external particles, i.e.\ $I,J = q, \bar{q}',\PW^{\pm}$, and the
universal (i.e.\ process-independent) functions
$G_{IJ}^{(\mathrm{sub})}(P_{IJ}, m_I, m_J)$ carry the respective
endpoint singularities. The charge flow of the external particle $J$ is
denoted by $\sigma_J$. Let us again state explicitly that the sum of the
one-loop virtual corrections and the endpoint contribution,
$\sigma_{q\bar{q}' \to V_1V_2}^{\mathrm{V}} + \sigma_{q\bar{q}' \to
  V_1V_2\gamma}^{\mathrm{E}}$, is IR finite by construction and defined
on a LO phase space solely parametrized by $\hat{s}$ and $\hat{t}$. In
our computation the quark masses are neglected whenever possible and
only introduced to regularize collinear singularities, whereas a finite
W mass is kept throughout the calculation. Therefore, the dipole
formulas for massive FS particles and massless IS particles have to be
applied here.
  
In case of a massless IS emitter $a$ and a massless IS spectator $b$ we find $P_{ab} =
\sqrt{\hat{s}}$ and
\begin{equation}
G^{(\mathrm{sub})}_{ab}(\hat{s},m_a^2) =
\mathcal{L}(\hat{s},m_a^2)-\frac{\pi^2}{3} + 2\,,
\end{equation} 
with the auxiliary function
\begin{equation}
\mathcal{L}(P^2,m^2) =
\ln\left(\frac{m^2}{P^2}\right)\ln\left(\frac{\lambda^2}{P^2}\right) +
\ln\left(\frac{\lambda^2}{P^2}\right) -
\frac{1}{2}\ln^2\left(\frac{m^2}{P^2}\right) + \frac{1}{2} \ln\left(\frac{m^2}{P^2}\right)\,.
\end{equation}
Here, soft singularities are regularized by an infinitesimal photon mass
$\lambda$, and a small quark mass $m_a$ is kept to regularize collinear
singularities, as mentioned before.  For the case of a massive FS
emitter and a massive FS spectator, which contributes to W-pair
production, $G^{(\mathrm{sub})}_{ij}(P_{ij}^2,m_i,m_j)$ is given by Eq.\ (4.10) from
Ref.~\cite{Dittmaier:1999mb}, with $P_{ij}^2 = \hat{s}$. The endpoint
contributions $G^{(\mathrm{sub})}_{ai}(P_{ia}^2,m_a,m_i)$ and $G^{(\mathrm{sub})}_{ia}(P_{ia}^2, m_i)$ for
massless IS emitters (spectators) and massive FS spectators (emitters)
can be found in Eq.\ (A.4) of the respective paper, where $m_a$ has been
set to zero in the expression for $G^{(\mathrm{sub})}_{ia}(P_{ia}^2,m_i)$. Here, the
auxiliary momentum is given by $P_{ia} = p_i - p_a$.

Note that the finite terms (i.e.\ terms not proportional $\ln(\lambda)$
or $\ln(m_q)$) of the endpoint contributions in Eq.~\eqref{eq:endpt} do
not have any physical meaning, since they are tailored to cancel certain
subtraction contributions from the real-radiation processes, which are
neglected anyway in the V+E approximation.  Nevertheless, our approach is
still justified since in the V+E approximation applied in this paper we
do not claim to control the QED part of the EW corrections, and, more
importantly, we have explicitly shown in Sect.~\ref{se:4ldetails} that
this approximation works remarkably well in $V$-pair production
processes. However, we clearly point out that this might be different
for different process classes, and the validity of the V+E approximation has
to be carefully checked if applied to other processes.
\end{appendix}

\subsection*{Acknowledgements}
This work was supported by the  
DFG Sonderforschungsbereich/Transregio 9 ``Compu\-tergest\"utzte 
Theoretische Teilchenphysik''.  SG acknowledges support from the EU
Initial Training Network ``MCnetITN'' and the Helmholtz Alliance
``Physics at the Terascale''. 

\nopagebreak
\nopagebreak

\end{document}